\documentclass[longauth]{aa}
\usepackage{amssymb,amsmath,amsfonts}
\usepackage{graphicx}
\usepackage{color}

\usepackage{xcolor}

\DeclareMathOperator{\e}{e}

\newcommand\eps{{\varepsilon} }

\newcommand\gO{{\cal O} }
\newcommand\cS{{S} }

\newcommand\cR{{R} }

\newcommand{\Ham}{\mathcal{H}}

\newcommand{\pp}{\left(p+1\right)}

\newcommand{\OO}{\mathcal{O}}
\newcommand{\G}{\mathcal{G}}

\newcommand{\er}{\xi_r}
\newcommand{\es}{\xi_s}

\definecolor{cadmiumgreen}{rgb}{0.0, 0.42, 0.24}

\newcommand{\orcid}[1]{\protect\href{https://orcid.org/#1}{\protect\includegraphics[width=8pt]{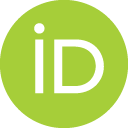}}}

\newcommand{\be}{\begin{equation}}
\newcommand{\ee}{\end{equation}}

\usepackage{natbib}

\usepackage{amsmath}
\usepackage{bbold}
\usepackage{tikz}
\usepackage{bm}

\usepackage[pdftex,colorlinks=true,linkcolor=blue,citecolor=blue,urlcolor=blue]{hyperref}

\begin{document}
\nolinenumbers
\title{Insights into planetary formation pathways from the \texttt{resonantstate} databases
%
}
\subtitle{
I. The resonant state of observed systems
}

\author{
A. Leleu$^{1}$\orcid{0000-0003-2051-7974},
Y. Eyholzer$^{1}$\orcid{0009-0004-3041-2249},
S. Hadden$^{2}$\orcid{0000-0002-1032-0783},
J. Couturier$^{1,3}$\orcid{0000-0001-8897-9978},
J.-B. Delisle$^{1}$\orcid{0000-0001-5844-9888},
R. Mardling$^{1,4}$\orcid{0000-0001-7362-3311},
D. C. Fabrycky$^{5}$\orcid{0000-0003-3750-0183},
D. Ragozzine$^{6}$,
E. Agol$^{7}$\orcid{0000-0002-0802-9145},
J.M. Almenara$^{1}$\orcid{0000-0003-3208-9815},
M. Goldberg$^{8}$\orcid{0000-0003-3868-3663},
J. Korth$^{1,9,10}$,
A. C. Petit$^{8}$\orcid{0000-0003-1970-1790},
M. Rosenqvist$^{1}$\orcid{0009-0009-6885-7450},
S. Marques$^{11,12}$\orcid{0009-0007-3825-7791},
K. Masuda$^{13}$\orcid{0000-0003-1298-9699},
G. Pichierri$^{14}$\orcid{0000-0003-3622-8712},
L. Borsato$^{15}$\orcid{0000-0003-0066-9268},
F. Dai$^{16}$\orcid{0000-0002-8958-0683},
I.-G. Jiang$^{17}$,
A.-S. Libert$^{18}$\orcid{0000-0003-2682-2793},
M. MacDonald$^{19}$\orcid{0000-0003-2372-1364},
S. Millholland$^{20,21}$,
S. Udry$^{1}$,
M.-T. Wang$^{22,23}$\orcid{0000-0003-3015-6455},
T. G. Wilson$^{24}$\orcid{0000-0001-8749-1962},
D. A. Yahalomi$^{21}$\orcid{0000-0003-4755-584X},
Y. Alibert$^{11,12}$,
A. Alnajjarine$^{25}$\orcid{0009-0005-5366-3371},
R. Burn$^{8}$\orcid{0000-0002-9020-7309},
A. C. M. Correia$^{25,26}$\orcid{0000-0002-8946-8579},
A. Emsenhuber$^{27}$\orcid{0000-0002-8811-1914},
A. Izidoro$^{28}$\orcid{0000-0003-1878-0634},
S. Shibata$^{29}$\orcid{0000-0002-5418-6336},
S. Ataiee$^{30}$\orcid{0000-0003-2594-3454},
R. Eltschinger$^{11,12}$\orcid{0009-0001-7351-3232},
Z.-C. Hu$^{31}$\orcid{0009-0000-6461-5256},
Y. Ishida$^{32}$\orcid{0009-0004-2879-1462},
E. Kokubo$^{33}$\orcid{0000-0002-5486-7828},
J. Laskar$^{25}$,
C. Mordasini$^{11,12}$\orcid{0009-0007-3825-7791},
D. Ségransan$^{1}$,
D. Tamayo$^{34}$\orcid{0000-0002-9908-8705},
L.-C. Yeh$^{35}$
}

\institute{
$^{1}$ Observatoire astronomique de l'Université de Genève, Chemin Pegasi 51, 1290 Versoix, Switzerland \\ 
$^{2}$ Canadian Institute for Theoretical Astrophysics, 60 St George St Toronto, ON M5S 3H8, Canada \\ 
$^{3}$ LTE, Observatoire de Paris, Univ. PSL, Sorbonne Univ., Univ. de Lille, LNE, CNRS, 61 Av. de l’Observatoire, 75014 Paris, France \\ 
$^{4}$ School of Physics \& Astronomy, Monash University, Clayton 3800, Australia \\ 
$^{5}$ Department of Astronomy \& Astrophysics, University of Chicago, 5640 S. Ellis Ave., Chicago, IL 60637, USA \\ 
$^{6}$ Brigham Young University Department of Physics \& Astronomy, N283 ESC, Brigham Young University, Provo, UT 84602, USA \\ 
$^{7}$ Astronomy Department and Virtual Planetary Laboratory, University of Washington, Seattle, WA 98195, USA \\ 
$^{8}$ Université Côte d’Azur, Observatoire de la Côte d’Azur, CNRS, Laboratoire Lagrange, UMR7293, Boulevard de l’Observatoire, 06304, Nice Cedex 4, France \\ 
$^{9}$ Instituto de Astrof\'{i}sica de Canarias (IAC), 38205 La Laguna, Tenerife, Spain \\ 
$^{10}$ Departamento de Astrof\'isica, Universidad de La Laguna (ULL), E-38206 La Laguna, Tenerife, Spain \\ 
$^{11}$ Space Research and Planetary Sciences, Physics Institute, University of Bern, Gesellschaftsstrasse 6, 3012 Bern, Switzerland \\ 
$^{12}$ Center for Space and Habitability, University of Bern, Gesellschaftsstrasse 6, 3012 Bern, Switzerland \\ 
$^{13}$ Department of Earth and Space Science, Graduate School of Science, Osaka University, Toyonaka, Osaka 560-0043, Japan \\ 
$^{14}$ Dipartimento di Fisica, Università degli Studi di Milano, Via Celoria 16, 20133 Milano, Italy \\ 
$^{15}$ INAF, Osservatorio Astronomico di Padova, Vicolo dell’Osservatorio 5, 35122 Padova, Italy \\ 
$^{16}$ Institute for Astronomy, University of Hawai`i, 2680 Woodlawn Dr, Honolulu, HI, 96822, USA \\ 
$^{17}$ Department of Physics and Institute of Astronomy, National Tsing Hua University, Hsinchu City 30013, Taiwan \\ 
$^{18}$ naXys, Departement of Mathematics, Rue de Bruxelles 61, 5000 Namur, Belgium \\ 
$^{19}$ Department of Physics, The College of New Jersey, 2000 Pennington Rd, Ewing, NJ 08628, USA \\ 
$^{20}$ Department of Physics, Massachusetts Institute of Technology, Cambridge, MA 02139, USA \\ 
$^{21}$ Kavli Institute for Astrophysics and Space Research, Massachusetts Institute of Technology, Cambridge, MA 02139, USA \\ 
$^{22}$ School of Astronomy and Space Science, Nanjing University, Nanjing 210023, China. \\ 
$^{23}$ Key Laboratory of Modern Astronomy and Astrophysics, Ministry of Education, Nanjing, 210023, People’s Republic of China \\ 
$^{24}$ Department of Physics, University of Warwick, Gibbet Hill Road, Coventry CV4 7AL, UK \\ 
$^{25}$ LTE, Observatoire de Paris, Univ. PSL, Sorbonne Univ., CNRS, 77 Av. Denfert-Rochereau, 75014 Paris, France \\ 
$^{26}$ CFisUC, Departamento de F\'isica, Universidade de Coimbra, 3004-516 Coimbra, Portugal \\ 
$^{27}$ Physics Institute, University of Bern, Sidlerstrasse 5, 3012 Bern, Switzerland \\ 
$^{28}$ Department of Earth, Environmental and Planetary Sciences, 6100 MS 126, Rice University, Houston, TX 77005, USA \\ 
$^{29}$ Department of Earth, Environmental and Planetary Sciences, MS 126, Rice University, Houston, TX 77005, USA \\ 
$^{30}$ Department of Physics, Faculty of Sciences, Ferdowsi University of Mashhad, Mashhad, 91775-1436, Iran \\ 
$^{31}$ Department of Astronomy, Tsinghua University, Beijing 10084, China \\ 
$^{32}$ Department of Astronomy, The University of Tokyo, 7-3-1 Hongo, Bunkyo-ku, Tokyo 113-0033, Japan \\ 
$^{33}$ Center for Computational Astrophysics, National Astronomical Observatory of Japan, Osawa, Mitaka, Tokyo 181-8588, Japan \\ 
$^{34}$ Department of Physics, Harvey Mudd College, 301 Platt Blvd, Claremont, CA 91711, USA \\ 
$^{35}$ Institute of Computational and Modeling Science, National Tsing Hua University, Hsinchu City 30013, Taiwan
}

 \abstract
  { 
Systems near mean motion resonances are especially informative to constrain the formation pathways of planetary systems, as their architecture depends sensitively on evolutionary history and can be precisely characterized via transit timing variations.
To compare observations and theories, we developed the two new \texttt{resonantstate} databases: an observational database which compiles analysis posteriors for hundreds of observed multi-planetary systems, and a simulation database that compiles the evolution of over ten thousand simulated systems which evolved via various physical processes (including accretion, planet-disk interactions, dynamical evolution, tidal dissipation).
In this study, we focus on observed pairs of planets near any first order mean motion resonance. In order to quantify the resonant state of such a pair, we employ the `Second Fundamental Model of Resonance' which projects the orbital elements and masses of a system onto three scale-free variables, two of which characterize its resonant state, independent of the specific first-order commensurability. 
In addition, TTV mass-eccentricity degeneracies are easily identified in these variables. We show that robust characterization of observed architectures---meaning free of degeneracies and independent of priors---is crucial to the emergence of trends for pairs near, but not inside, a resonance.

We find that both the close-in ($\smash{P_{\rm in}\!<\!5}$d) and far-out ($\smash{P_{\rm in}\!>\!5}$d) sub-populations have few resonant pairs, and more than twice as many pairs just wide of resonance as narrow of it. Far-out pairs have moderate eccentricities, similar between the resonant and near-resonant pairs; while close-in pairs have small eccentricities.
In addition, close-in pairs have a larger departure from exact commensurability. Our findings are consistent with a similar formation history of both sub-populations, followed by a slow tidal evolution of the close-in population. If the bulk of the close-in population was in resonance at the end of the proto-planetary disk, our results imply that most had to break away from resonance before tides played a significant role. In order to reinforce these results and broaden the scope of the observational database, we welcome new additions of posteriors of dynamical analysis of multi-planetary systems. 
}


\titlerunning{Resonant state of observed pairs}

\authorrunning{Leleu et al.}

\maketitle
\nolinenumbers

\section{Introduction}

The observed architectures of planetary systems are the byproducts of numerous physical processes which occur during their formation in the proto-planetary disk, as well as millions to billions of years of subsequent evolution after disk dispersal.
Among the possible architectures, systems hosting pairs of planets whose period ratio is close to a low-order commensurability are of particular interest. 
{Such structures form naturally when the difference in disk torques balances the planet-planet torque during slow convergent migration in the proto-planetary disk, a state made possible because dissipation tends to damp the eccentricities to small (but non-zero) values \citep{GoTre1979,Weidenschilling1985,NePa2003,Masset2006,BatPet2023}.} Moreover, the force between planets at conjunction tends to be in the same direction over many orbits (typically hundreds for pairs of sub-Neptunes), resulting in cumulative changes in the orbital periods, eccentricities and orientations which in turn are detectable as transit timing variations (TTVs) \citep{Agol2005,NeVo2016,AgolFabrycky2025}. 

The discovery of hundreds of transiting multi-planetary systems by the Kepler mission has enabled  statistical comparisons between observations and theory.
Trends related to (nearly-)resonant systems were discovered in the distribution of period ratios \citep{Fabrycky2014,Delisle2014letter},
but also other quantities such as the planetary densities and mutual inclinations \citep{Millholland2020,Millholland2022,Leleu2024b}, and stellar age \citep[][using galactic velocity dispersion and stellar clusters, respectively]{Hamer2024,Dai2024}.
The origin of some features, such as the overdensity of pairs about one percent above the exact commensurability, is still discussed \citep[e.g.][]{Lee2013,Fabrycky2014,Delisle2014,Choksi2020,GoBa2023}.

TTVs have emerged as the most sensitive probe of small planet masses and orbits and are available in many systems where radial velocity measurements are infeasible. {TTVs can be analysed by fitting transit timings that were previously extracted from the lightcurve} \citep[e.g.][]{Rowe2014,Holczer2016}, or by photodynamical analysis that directly model the light curve while taking into account planet-planet gravitational interactions \citep[][]{RaHo2010,Carter2012}. 
In principle, TTVs encode information about the planet-to-star mass ratios and all orbital elements.\footnote{Although only the difference in ascending node longitudes.} In practice, finite timing precision and observational baseline
Usually result in degeneracies between the eccentricities and their associated longitudes of periastron (\S~\ref{sec:degen}), and worse still, between these quantities and the masses themselves
\citep[see also \S~\ref{sec:degen}]{HaLi2016,NeVo2016,Jontof2016,Mills2016,Agol2021,Leleu2023,Almenara2025}. Often, the mass-ratios of pairs of planets are better constrained than the ratio of the planets to the star \citep{Holman2010, Jontof2016}.
In addition, while a Bayesian photodynamical analysis contains information about these degeneracies, it is common that only median parameter values and associated error bars are reported, leading to a misleading view of the complete solution in the full parameter space consistent with the data \citep{Leleu2023}. 

To alleviate this situation, samples of the posterior of the TTV analysis can be used to analyze the dynamical state of a system; these samples are sets of masses and orbital elements at epoch that explain the data. {Any quantity of interest can be computed for each set of orbital elements, preserving their correlations}. This has been done to analyze the resonant state of individual systems \citep[e.g.][]{Nesvorny2022,Leleu2022}. In \cite{HaLi2017}, {the authors analysed pre-extracted TTVs for 145 Kepler planets, enabling a statistical comparison of their eccentricities}. The samples of their TTV analysis were then used by \cite{GoBa2023} to compare with simulations of pairs of Super-Earths that underwent specific physical processes that could shape planetary systems.

Constraining processes of formation and evolution of planetary systems from observed architectures was one of the main goals of the Geneva Resonant State Workshop\footnote{\url{https://indico.global/event/13953/}}, which the authors attended in early June of 2025. To enable comparison of observations with theory at a larger scale, we started the observation and simulation \texttt{resonantstate} databases, which are presented in \S\ref{sec:databases}. In this first paper, we focus on the observed population. In \S\ref{sec:framework}, we describe the framework we use to quantify the resonant state of pairs of planets. In \S\ref{sec:obs}, we present the resonant state of the observed population, and describe new observational trends. We then discuss our results in \S\ref{sec:conclusion}.

\section{The databases}
\label{sec:databases}

To facilitate the comparison of the architecture of observed and simulated systems, we created two databases in which the osculating astrocentric orbital elements, planetary masses, and other relevant quantities of both observed and simulated systems are stored in the same format. The databases can be explored using the \texttt{resonantstate} package\footnote{\url{https://github.com/adleleu/resonantstate}}. The folder `paper1' contains notebooks to reproduce figures of this paper and enable users analyze their own systems. 
The structure of the database is given in appendix \ref{ap:DB}.

\begin{figure*}
\begin{center}
\includegraphics[width=0.99\linewidth]{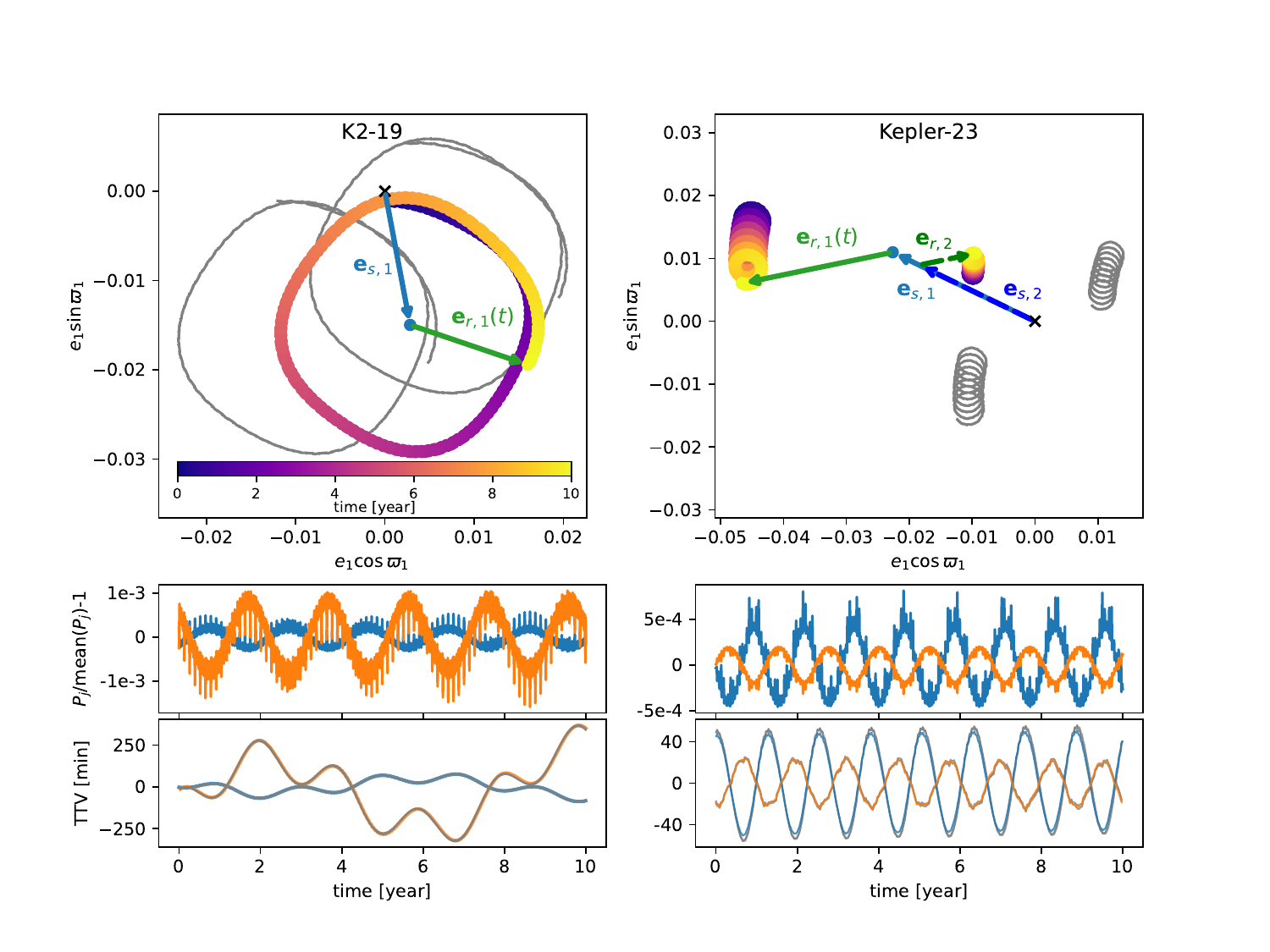}
\caption{\label{fig:elements}{ Examples of evolution of orbital elements for a set of initial conditions for K2-19 \citep[][left]{Almenara2025} and Kepler-23 \citep[][right]{Leleu2023}.}
The top panels show the temporal evolution of $e_1 \cos \varpi_1$ and $e_1 \sin \varpi_1$.
The middle panels show the temporal evolution of the orbital periods of the planets, and the bottom panels show a forward model of the TTVs. 
The grey curves in the top and bottom panels show two other sets of initial conditions that are consistent with the data. For Kepler-23, the outer planet's eccentricity vector is also shown in the upper panel. 
}
\label{fig:elements} 
\end{center}
\end{figure*}

The observation database stores posterior samples from the MCMC analysis of observational data, as well as a metadata table\footnote{\url{https://dace.unige.ch/resonantStateObservationsCatalog/}} which summarizes high-level information on each entry and facilitates their selection for in-depth analysis. The samples are hundreds or thousands of sets of orbital elements and planetary and stellar parameters (notably masses), which are compatible with observational data. The rate at which points in the parameter space are included in the posterior samples is proportional to their posterior probability. {Figure \ref{fig:elements} shows  three sets of initial conditions for each of the systems K2-19 and Kepler-23. Each configuration can be integrated to analyse its evolution over time, or to predict future transits. The time-dependent orbital elements can also be used to compute dynamically-relevant quantities while preserving the correlation between parameters, as we will see in \S \ref{sec:framework}}.

The version of the database used for this paper has 1714 planets in 675 individual systems, contributed by 13 teams \citep{2015MNRAS.453.2644A,2016A&A...595L...5A,MacDonald2016,Mills2016,HaLi2017,2018A&A...615A..90A,2018MNRAS.478..460A,2021A&A...655A..66L,Agol2021,Leleu2022,2022A&A...663A.134A,Leleu2023,2023A&A...675A.115K,2023AJ....165...33D,2023NatAs...7..206P,Leleu2024,2024A&A...689A..52B,2024AJ....168...31W,Masuda2024,2024ApJ...971L..28K,2025PSJ.....6..267L,2025AJ....169..243J,Leleu2026,Ragozzine2026}, mostly from TTV analysis, with a few combined TTV and RV analysis, and currently a single system characterized by RV alone \citep[GJ 876][]{Millholland2018}.
Some of these systems have been analyzed more than once by one or more teams, with various models. The metadata table summarizes this information, and provides a `default' flag that indicates the preferred solution of each author. In \S\ref{sec:selection}, we give our selection criteria for this paper.

The second database stores the time evolution of simulations of the formation and evolution of planetary systems under various physical processes: 2D hydrodynamical simulations \citep{Ataiee2021}, pairs of planets evolving in a laminar, turbulent, planetesimal 1D disk or tides \citep{GoBa2023}, population synthesis runs from the Bern model \citep{Emsenhuber2021} and `Breaking-the-chain' runs \citep{Izidoro2017,Shibata2025}, as well as N-body/collisional evolution \citep{Kokubo+2025}. We present this database further in the second paper of the series.

\section{Framework for the study of resonant and near-resonant pairs}
\label{sec:framework}

\begin{figure}
    \centering
    \includegraphics[width=.99\linewidth]{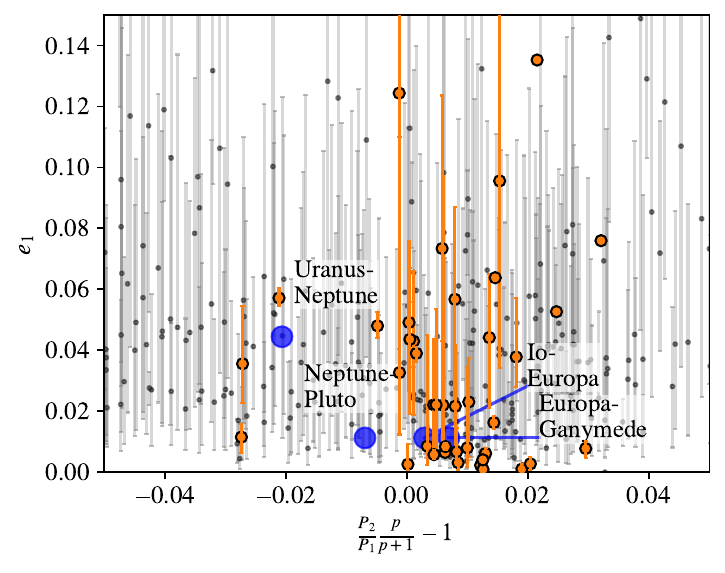}
    \caption{Posteriors of near-resonant pairs from the observation database, following the selection described in \S \ref{sec:selection}. Orange shows pairs of planets that are robust to the mass-eccentricity degeneracy \citep{HaLi2017,Leleu2023}, or whose mass precision is better than 7\% (see \S \ref{sec:degen}).
    Blue circles show pairs from the solar system.
    \label{fig:Pve1}}
\end{figure}

\begin{figure}
    \centering
    \includegraphics[width=.99\linewidth]{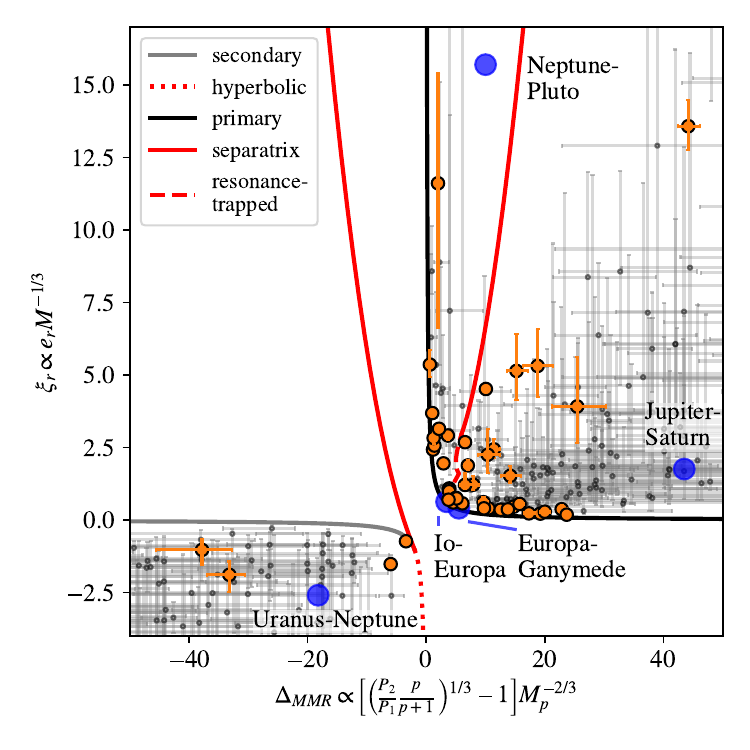}
    \caption{Projection of the posteriors of near-resonant pairs from the observation database on the rescaled quantities of the second fundamental model that we describe in \S \ref{sec:SFM}. This projection enables the description of the resonant state of a pair, e.g. its position with respect to the equilibria and separatrix of the phase space. The posteriors are the same as Fig. \ref{fig:Pve1}.  }
    \label{fig:SFMobs}
\end{figure}

In this study we focus on the orbital architecture of pairs of planets near first order resonances ($P_2/P_1\sim(p+1)/p$) for two planets with periods $P_1$ and $P_2$, and an integer $p$. The distance to commensurability in period ratio and the planets' eccentricities may at first appear to be a natural choice to describe these architectures. However, the projection of observed systems in that plane is difficult to interpret, as is apparent in Fig. \ref{fig:Pve1}. The reasons for this are threefold: firstly, part of each planet's eccentricity vector is generally not constrained by TTVs, which leads to large uncertainties on the full eccentricity vectors. Secondly, the position of a pair in the resonant phase space depends on 10 parameters ($2\times 4$ orbital elements and two planet-to-star mass ratios). Thirdly, the position of the features of the resonant phase space, such as its equilibria and resonance boundaries, depends on the orbital elements and masses of the system under consideration: two pairs could occupy the same position in the plane shown in Fig \ref{fig:Pve1}, when in fact one sits at the  center of the resonance while the other is outside of the resonant boundaries. 

In this section we show how to project the state of a pair of planets near a first-order MMR onto the plane of the Second Fundamental Model of Resonance \citep[SFMR,][see Fig. \ref{fig:SFMobs}]{HenLe1983}, allowing one to distinguish between systems with different resonant states and hence formation histories. In particular, the SFMR employs the fact that the time-varying {\it resonant} parts of the eccentricity vectors, which dominate the observed TTVs, are associated with energy and angular momentum exchange between the orbits and are proportional to a linear combination of both vectors, while the remaining {\it secular} parts, which are constant at first-order in eccentricity and slowly varying at second order, are proportional to a different and independent linear combination of the eccentricity vectors.\footnote{That the secular parts only vary at second order is the reason that TTVs generally contain very little information about these components.} This is the {\it rotation transformation} discovered by \cite{SeFe1984}, which allows the SFMR of \cite{HenLe1983} to be generalised from the restricted problem to that of two massive planets. In particular, it reduces the full coplanar problem to one which is integrable, thereby linking all 10 system parameters to a single 1D curve, whose position in phase space is determined by the total angular momentum deficit and the energy associated with resonant oscillations. The latter are characterized by the two dimensionless parameters $\delta$ and $\xi_r$, which correspond to the $x$ and $y$ axes of the SFMR plane (see top-left panel of Figure~\ref{fig:SFM_anarea} for this standard representation). States corresponding to separatrices and other boundary curves can also be represented in the SFMR plane, so that the position of a specific system relative to such curves provides direct information about its resonant state, regardless of masses and particular first order MMR. The gain of the projection on the SFMR can also be illustrated by the position of pairs from the solar system (Fig. \ref{fig:Pve1} and \ref{fig:SFMobs}).  For example, for the chosen date ($2451545.0$JD), the Neptune-Pluto pair does not necessarily appear to be in resonance in Fig. \ref{fig:Pve1}, while the pair is clearly within the resonance boundary in Fig. \ref{fig:SFMobs}. In term of the strength of neighboring 2:1 MMR, Jupiter-Saturn ($P_{Sat}/P_{Jup}\!\approx\!2.48$) is only twice as far as Uranus-Neptune ($P_{Nep}/P_{Ura}\!\approx\!1.95$).

Projection onto the SFMR plane has been used to study the resonant dynamics associated with various physical processes \citep[e.g.][]{Delisle2012,Batygin2015}, and to discuss the resonant state of individual systems \citep[e.g.][]{BaDeHo2015,Panichi2019,Petit2020,Leleu2022}. In this study we use this powerful technique to compare the resonant states of a broad range of observed and simulated systems. In the limit of coplanarity, we will introduce three scale-free parameters that are well suited for meaningful comparisons between evolution models and observations: a proximity to resonance $\Delta_{\rm MMR}$ (eq.\ \ref{eq:DMMR}), and the (rescaled) resonant and secular eccentricities $\er$ and $\es$ (sect.\ \ref{sec:SFM}).


\subsection{Examples of motion}
\label{sec:examples}
We start by describing the examples of motion seen in Fig. \ref{fig:elements}. To describe the orbits, we use the heliocentric canonical set of orbital elements of \citep{LaRo1995}. For a pair of planets near, or within, a first order MMR, their eccentricity vectors can be decomposed as \citep[Appendix \ref{ap:ejwj};][]{SeFe1984,Henrard1986}:
\begin{equation}
\begin{aligned}
e_1(t) e^{i\varpi_1(t)} &=& &e_{r,1}(t) \e^{i \varpi_r(t)}
&+  \ \ e_{s,1} \e^{i \varpi_s}\, , \\
e_2(t) e^{i\varpi_2(t)} &=& - \ \ &e_{r,2}(t) \e^{i \varpi_r(t)}
&+ \ \  e_{s,2} \e^{i \varpi_s},
\end{aligned}
\label{eq:esplit}
\end{equation}
where $\smash{e_j(t) \e^{i\varpi_j(t)}}$, $\smash{e_{r,j}(t) \e^{i\varpi_r(t)}}$ and $\smash{e_{s,j} \e^{i\varpi_s}}$ are the complex representations of the Runge-Lenz vector of planet $j$, $\vec{e}_j(t)$, and its resonant $\vec{e}_{r,j}(t)$ and secular $\vec{e}_{s,j}$ components. The secular components are constant (at first order in eccentricity), and are aligned and of similar magnitude, while the time-varying resonant components are anti-aligned with an amplitude ratio proportional to the planet mass ratio; see Appendix \ref{ap:ejwj}.

In our model, $\vec{e}_{s,j}$ (in blue in Fig \ref{fig:elements}) is constant, although a slow evolution can be seen in the numerical simulation in the top-right panel of Fig \ref{fig:elements}. In contrast, $\vec{e}_{r,j}(t)$ evolves over the resonant period, $P_{res}$. We denote $\smash{\eps=((m_1+m_2)/m_\star)}$. Inside the resonance (left panels of Fig.\,\ref{fig:elements}), $P_{res} \propto P_{orb} \eps^{-2/3}$ ($\approx 2$ years for K2-19), leading to the oscillation of amplitude and phase of the green vector in Fig \ref{fig:elements}. The planets' semi-major axes, hence orbital periods, evolve on the same timescale: the resonant motion is responsible for energy exchange between the two orbits. In addition, $\vec{e}_{r,1}$ precesses at the mean super frequency
$\overline{f}_{super}$, which is the frequency of precession of the line of conjunction \citep{Agol2005,Lithwick2012}:
\begin{equation}
\begin{aligned}
\overline{f}_{super}=2\pi \left( \frac{p}{\overline{P_1}}-\frac{p+1}{\overline{P_2}} \right)\, ,
\label{eq:supa}
\end{aligned}
\end{equation}
where $\overline{P_j}$ is the orbital period of planet $j$ averaged over the resonant motion. The associated period, called the ``super period,'' is $\approx 8$ years for K2-19. 

Inside the resonance the resonant timescale is often well-separated from the super period, while outside of the resonance (right panels of Fig \ref{fig:elements}), the resonant period becomes close to the super-period resulting in a single periodicity in the evolution of the eccentricity vectors and orbital periods on short baselines. In this case $\vec{e}_{r,j}(t)$ has a constant component and a small oscillation at the super period. 

In this study, we focus on the resonant motion responsible for the evolution of $\vec{e}_{r,j}$ and $P_j$. The splitting of the components of the eccentricities (Eq.\ \ref{eq:esplit}) is an intermediary step toward the SFMR, but additional transformations are required to benefit from further dimensionality reduction and the scale-invariance of the SFMR. These transformations, which rely on the work of \cite{SeFe1984} and \cite{Henrard1986}, are derived in Appendix \ref{ap:SFM}. In the next section, we review some of the SFMR properties and describe the associated motion.

\subsection{The second fundamental model of resonance}
\label{sec:SFM}

A set of orbital elements at epoch for a pair of planets near, or within, any first-order mean-motion resonance can always be converted into the scale-free variables of the second fundamental model of resonnance (SFMR), which
enables the computation of its position in the resonant phase space, determines its evolution, and allows us to compare its resonant state to other pairs of observed or simulated planets. 

\begin{figure*}
\begin{center}
\includegraphics[width=0.39\linewidth]{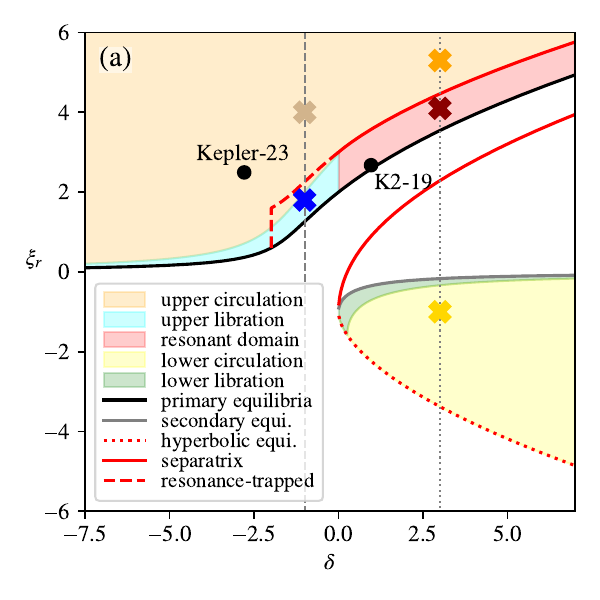}
\includegraphics[width=0.39\linewidth]{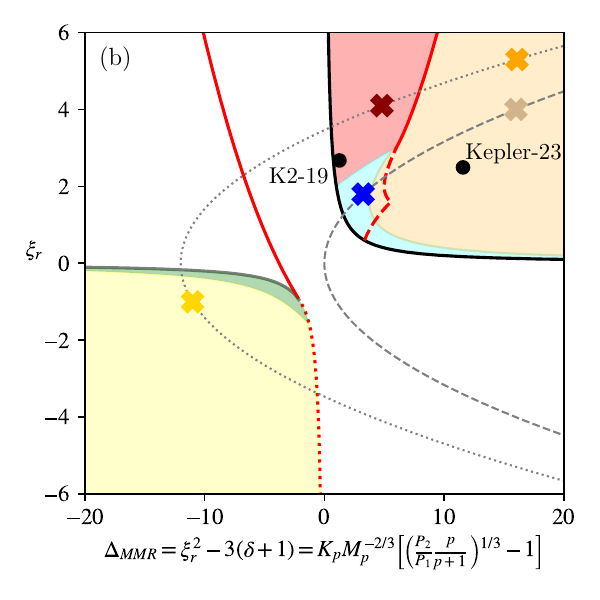}
\includegraphics[width=0.39\linewidth]{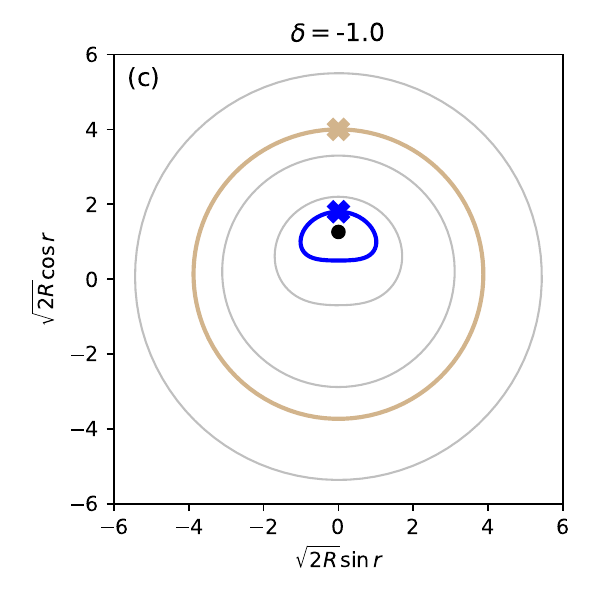}
\includegraphics[width=0.39\linewidth]{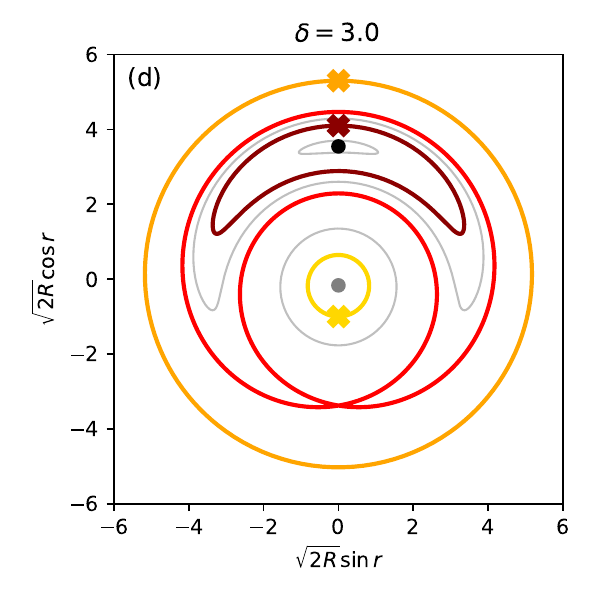}
\caption{Top: phase space of the SFMR; (a) in the ($\delta, \er$) plane, and (b) in the ($\Delta_{\rm MMR}$, $\er$) plane.
Curves of constant $\delta$ are shown in grey. For $\delta=-1$ (dashed), the phase space in the ($R,r$) variables is shown in panel (c), while $\delta=3$ (dotted) is shown in panel (d).  
Trajectories follow energy curves such as the grey and colored examples. The colored examples also show with $\times$ the phase at which they cross the ($\delta,\er$) plane with maximum value of $|R|$ and are represented in the upper panels in the coloured area (white areas show the place where energy curves cross this plane a second time and are not used). 
Colored areas of the upper panels describe various motions. Red is the `formal' resonant domain. The term `libration' refers to an area in which the resonant angle $r$ oscillates around a given value with semi-amplitude smaller than $\pi$, while `circulation' implies that $r$ evolves monotonically and explores all possible values. The label `upper' (resp. `lower') refers to orbits above (resp. below) the exact commensurability. 
The positions of our two examples of Fig.~\ref{fig:elements} are shown by black circles in the top panels.
}
\label{fig:SFM_anarea} 
\end{center}
\end{figure*}

When describing the motion of a pair near-resonant planets, the eccentricity vector components, $e_{r,j}$ and $e_{s,j}$ of eq.\ \ref{eq:esplit}, are proportional to $\smash{\sqrt{2R}}$ and $\smash{\sqrt{2S}}$, respectively, with:
\begin{equation}
\begin{aligned}
\cR / K_r &= f^2_1 e^2_1 + f_2^2 e_2^2 + 2 f_1 f_2 e_1 e_2 \cos (\varpi_1 - \varpi_2) \\
\cS / K_s  &=  f_2^2 \frac{\Lambda_{1,0}}{\Lambda_{2,0}} e_1^2 +  f^2_1 \frac{\Lambda_{2,0}}{\Lambda_{1,0}} e^2_2 - 2 f_1 f_2 e_1 e_2 \cos (\varpi_1 - \varpi_2)\, ,
\end{aligned}
\label{eq:RS}
\end{equation}
where $K_r$, $K_s$, $\Lambda_{1,0}$ and $\Lambda_{2,0}$ depend on the planets’ orbital periods and masses (Eqs.\ \ref{eq:poincare}, \ref{eq:RpSp} and \ref{eq:Komega}), $f_1$ and $f_2$ depend on $\alpha \equiv a_1/a_2$ (eq. \ref{eq:f1f2}), while $\varpi_r$ and $ \varpi_s$ additionally depend on the eccentricity vectors. $R$ and $S$ are canonical momenta, and their associated angles are:
\begin{equation}
\begin{aligned}
r &= -p \lambda_1 + (p+1) \lambda_2 - \varpi_r \\
s  &=  -p \lambda_1 +(p+1) \lambda_2 - \varpi_s  \, ,
\end{aligned}
\label{eq:RS}
\end{equation}
where $\lambda_j$ is the mean longitude of planet $j$.

The quantity $S$ is constant in this model. The resonant motion is therefore determined by the evolution of $(R,r)$, which is described by the one degree-of-freedom SFMR \citep[][]{HenLe1983}. Examples of the evolution of variables $(R,r)$ in that model are given in Fig.~\ref{fig:SFM_anarea}. These trajectories depend on the value of the constant parameter $\delta$ which is linked to the angular momentum deficit of the pair (eq. \ref{eq:delfsup}) :
\begin{equation}
\begin{aligned}
\delta \approx  \frac{2}{3} \left( \frac{C_\delta f_{super}}{M^{2/3}} + R \right) -1\, ,
\label{eq:delt}
\end{aligned}
\end{equation}
where $C_\delta$ depends on the averaged orbital periods of the planets, and $\smash{M \approx \eps = (m_1+m_2)/m_\star}$ for large $p$
(eq. \ref{eq:MMC}). Since $\delta$ is constant, Eq.\ \ref{eq:delt} links the evolution of the eccentricities and the periods through $R$ and the instantaneous super frequency : 
\begin{equation}
\begin{aligned}
{f}_{super}=2\pi \left( \frac{p}{P_1}-\frac{p+1}{P_2} \right)\, .
\label{eq:superfreq}
\end{aligned}
\end{equation}

{Depending on $\delta$, the phase space is similar to a pendulum ($\delta\! >\!>\! 0 $) or a harmonic oscillator ($\delta\! <\!< \!0 $). For all values of $\delta$, an elliptic fixed point ($ell_1$) exists, as shown in black in Fig.~\ref{fig:SFM_anarea} (a). These fixed points, associated with orbits that are either resonant or above the exact commensurability, will be called the primary fixed points}. For ($\delta \ge 0 $), there exist also (i) an additional elliptic fixed point ($ell_2$), associated with period ratios below the exact commensurability, that we call the secondary fixed point; and (ii) a hyperbolic equilibrium and a separatrix that emanate from it, shown as the red curve in Fig.~\ref{fig:SFM_anarea} (a).
The variables ($R$,$r$) evolve along the energy curves in grey in Fig.~\ref{fig:SFM_anarea} (c) and (d).
{A set of initial conditions is said to be resonant if ($\delta \ge 0 $) and its trajectory lies within the separatrix.}

{For a given value of $\delta$, all energy curves, hence all trajectories, intersect the plane $\sin r=0$, and all trajectories are periodic}. We will therefore forfeit the information on the position of the initial condition of the pair in the ($\sqrt{2R}\cos r$,$\sqrt{2R}\sin r$) plane at a given epoch and instead systematically follow the energy curve to the maximum value of $R$. This is reached for $r= 0$ for pairs above the exact commensurability, and for $r= \pi$ for pairs below the exact commensurability. We denote $\er$ the value of $\sqrt{2R}\cos r$ at that point
and refer to it as the rescaled resonant eccentricity of the pair. 
By analogy, we define the rescaled secular eccentricity of the pair $\es = \sqrt{2 \cS}$.

\begin{table}[!h]
\centering
\caption{$K_p$ used to compute $\Delta_{\rm MMR}$ and $\Delta_{p+1:p}$.}\label{tab:Kp}
\begin{tabular}{c|c c c c c c}
    p & $1$ & $2$   & $3$ & $4$ & $5$ & $6$ \\
    \hline
$K_p$ & $5.64$ & $3.56$ & $3.18$ & $2.92$ & $2.73$ & $2.58$
\end{tabular}
\end{table}

To represent the ($\delta,\er$) plane in a way that might be more intuitive, we follow \cite{Lin2025} and replace the abscissa by:
\begin{equation}
\begin{aligned}
\Delta_{\rm MMR} &= \er^2 - 3(\delta+1) = K_p M_p^{-2/3} \left[ \left(\frac{P_2}{P_1} \frac{p}{p+1} \right)^{1/3} -1 
\right], \\
 &\propto M_p^{-2/3} \overline{P}_1 f_{super} 
\label{eq:DMMR}
\end{aligned}
\end{equation}
where $M_p$, defined in eq (\ref{eq:Mp}), is equal to $\smash{\eps=(m_1+m_2)/m_\star}$ when $m_1\!=\!m_2$, and the $K_p$ are constants that depend only on the chosen $p+1\!:\!p$ MMR (see eq. \ref{eq:DMMRap} and Table \ref{tab:Kp}). In equation (\ref{eq:DMMR}), the instantaneous periods, and super frequencies, are the value reached when $\sin(r)=0$,a s discussed in the previous paragraph. This quantity, which we denote the proximity to resonance, is reminiscent of a similar parameter used in other TTV studies \citep[e.g.][]{Lithwick2012,HaLi2014}, but is formally distinct, as it incorporates a dependence on planetary masses and $p$ in order to identify dynamically equivalent states. Using $\Delta_{\rm MMR}$ places pairs below commensurability on the left hand-side of the plot and above commensurability on the right, with a distance to $0$ proportional to the super-frequency, and the $\er$ that go to infinity toward exact commensurability; see Fig.~\ref{fig:SFM_anarea} (b). In this representation, the points for which the trajectories intersect the $\sin(r)=0$ plane land on a parabola (constant $\delta$), as shown by the dashed/dotted grey lines in Fig. \ref{fig:SFM_anarea} (b).  

In the ($\Delta_{\rm MMR},\er$) plane (and equivalently in the $\delta,\er$ plane) we define the resonant area in red, the upper (resp. lower) circulation where $r$ is circulating and $P_2/P_1>(p+1)/p$ (resp. $P_2/P_1<(p+1)/p$). Similarly, we define the upper and lower libration areas when $r$ is librating between bounded values. 
In order to discuss the distance of a pair of planets to the relevant equilibria for a given value of $\delta$, we write:
\begin{equation}
 \Delta \er = \er-\xi_{ell_k}(\delta),
\end{equation}
where $\xi_{ell_k}(\delta)$ is the value of $\er$ at the elliptic point ($k=1$ for pairs in resonance or $P_2/P_1>(p+1)/p$, and $k=2$ otherwise) following the curve of constant $\delta$ of the pair, and $\Delta \er$ the distance to that equilibrium.
$\Delta \er=1$ when the separatrix appears for $\delta=0$.  
We note that the dynamics of a pair of planets near the primary equilibrium does not change qualitatively at $\delta=0$, and that the dissipative processes that can form resonant pairs will lead to an equilibrium eccentricity value independent to the value at which the separatrix appears \citep[e.g.][]{LePe2002,Delisle2015}. Pairs can therefore be trapped in resonance before they enter the formally resonant domain. This motivates the definition of the resonance-trapped domain, which adds to the formally resonant domain an extension with $\Delta \er=1$ down to $\delta=-2$ (red dashed curves in Fig. \ref{fig:SFM_anarea} a and b). This lower bound on $\delta$ is arbitrary, and corresponds to a primary equilibrium with a non-rescaled eccentricity of $e_{r,1}\!\sim\! 0.005$ for a pair of $10M_\oplus$ near a 3:2 MMR. 

\subsection{Pair selection} 
\label{sec:choice}

\begin{figure}
\begin{center}
\includegraphics[width=0.99\linewidth]{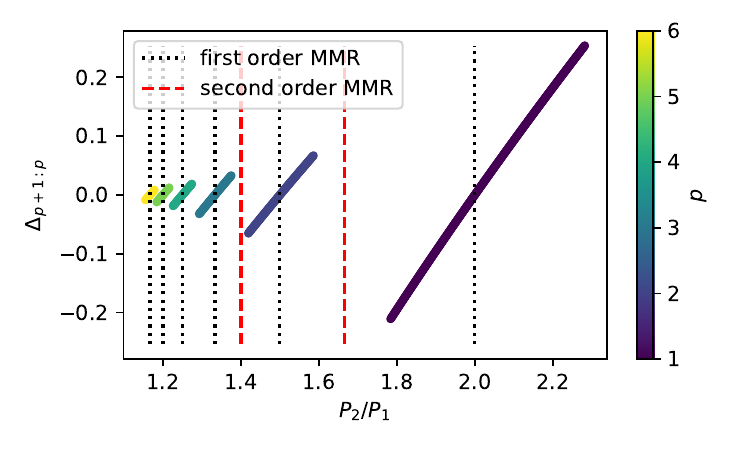}
\caption{Distance to commensurability $\Delta_{p+1:p}$ as function of period ratio. The color indicates to which resonance $p$ a pair is assigned. Pairs with period ratio between the displayed ranges are not used.
}
\label{fig:Dpp1p} 
\end{center}
\end{figure}

In order to use the SFMR for a given planet pair, one needs to first choose a specific $p+1\!:\!p$ resonance from their period ratio.
For this, we define a distance to MMR $\Delta_{p+1:p}$ that takes into account the strength of each $p+1\!:\!p$ resonance. 
Given the expression of $\Delta_{\rm MMR}$, we choose:
\begin{equation}
\begin{aligned}
\Delta_{p+1:p}=M_p^{2/3}\Delta_{\rm MMR} = K_p  \left[ \left(\frac{P_2}{P_1} \frac{p}{p+1} \right)^{1/3} -1 
\right]. 
\label{eq:Dpp1p}
\end{aligned}
\end{equation}
This distance allows us to choose which resonance is the closest in terms of ($\Delta_{\rm MMR}$,$\er$), regardless of the planetary masses, since the width of all first order MMRs scales as $M_p^{2/3}$ which only weakly depends on $p$. Since the first order MMRs become more closely spaced in period ratio as $p$ increases, this distance will map a given resonance to a wider range of period ratios above commensurability than below. To avoid this, we assign a pair of planets with a distance $\Delta_{p+1:p}$ to a given resonance $p$ only if that resonance is still the closest if its period ratio is given by $-\Delta_{p+1:p}$ (this ensures a symmetric range of distances that is associated with a given $p$, see Fig. \ref{fig:Dpp1p}). Otherwise, the pair is not {used in the following sections}. This also conveniently removes any pairs near the second order MMRs 5:3 and 7:5, and provides an upper limit to the range of considered period ratios.

\subsection{Applicability of the model }
\label{sec:validity}

Projecting onto the SFMR the orbital elements at epoch and masses of a pair of planets embedded in a 2+ planetary system implies several assumptions. First, the projection is valid for small eccentricities (first order in the perturbation; see Appendix \ref{ap:SFM}), small mutual inclinations ($\lesssim 10^\circ$), and timescales of a few tens of years for sub-Neptunes with orbital periods of tens of days ($\es$ is assumed constant).

Second, the input orbital elements are assumed to be the mean (averaged over fast frequencies, such as orbital motion) heliocentric canonical coordinates \citep[astrocentric positions, barycentric velocities][]{LaRo1995}. Using the non-averaged astrocentric variables of the database a priori induces errors of the order of $\eps = (m_1+m_2)/m_\star$ in the SFMR parameters, hence much smaller than the precision of the observations. 
In a few cases (large $\eps$, large $p$) this effect was found to shift pairs by a few tenths of $\Delta \xi_r$ in the SFMR (see the example\footnote{\url{https://github.com/adleleu/resonantstate/blob/main/paper1/osculating_to_mean_elements.ipynb}} of Kepler-307), but we verified these to be rare enough to not affect the trends discussed in the rest of the paper.

Third, neighboring planets affect the orbital elements of the considered pair. We experimented on their effect on $\Delta_{\rm MMR}$ and $\er$ by varying the mass of the perturber. Using various tests presented in Appendix \ref{ap:validity}, we found that the effect on $\er$ was limited below $0.2$ if the mass of the perturber is smaller than half the sum of the masses of the considered pair, or a quarter if either planet is involved in non-consecutive first-order MMRs. We will therefore discard pairs whose neighbors are above this threshold for sections \ref{sec:Pmin}-\ref{sec:upperDer}.

\subsection{TTV analysis and degeneracies}
\label{sec:degen}

\begin{figure}
\begin{center}
\includegraphics[width=0.99\linewidth]{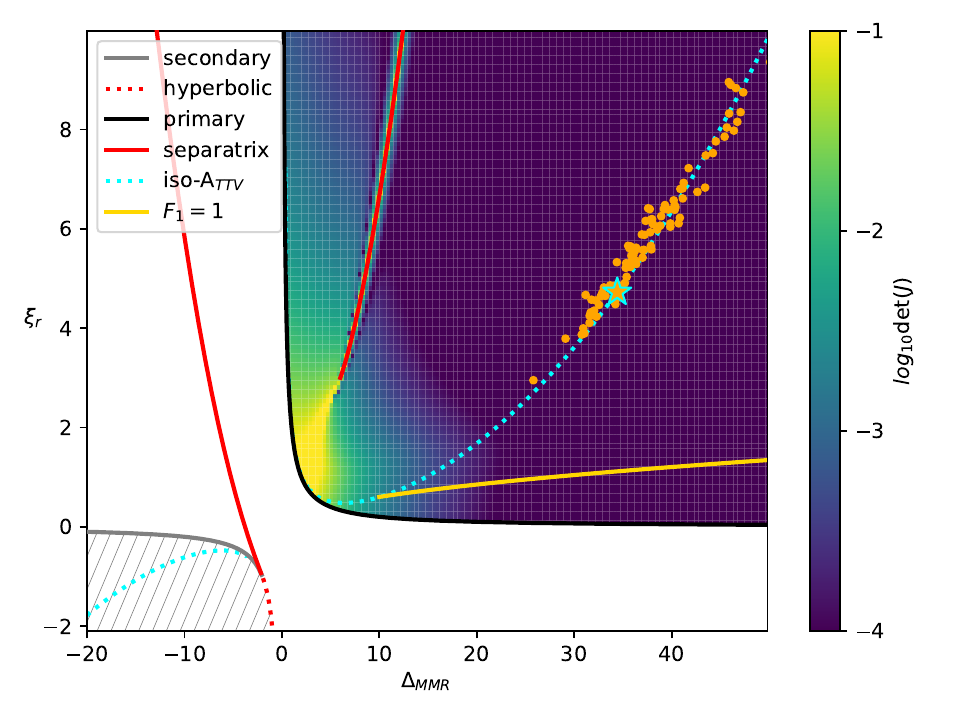}
\caption{Degeneracy in the SFMR for sinusoidal TTVs in the case of Kepler-52, with $P_2/P_1=2.079$. The dashed area shows the part of the phase space that cannot be reached with this period ratio. The color indicates the amount of degeneracy, with darker color implying stronger degeneracy, and is independent of the considered pair. The orange dots are samples from the posterior of Kepler-52 b and c. The cyan curve indicates all points of the plane that have the same value of $A_{TTV}$ as the star-shaped sample (independent of $M$). 
}
\label{fig:SFM_degen} 
\end{center}
\end{figure}

 \begin{figure}
\begin{center}
\includegraphics[width=0.99\linewidth]{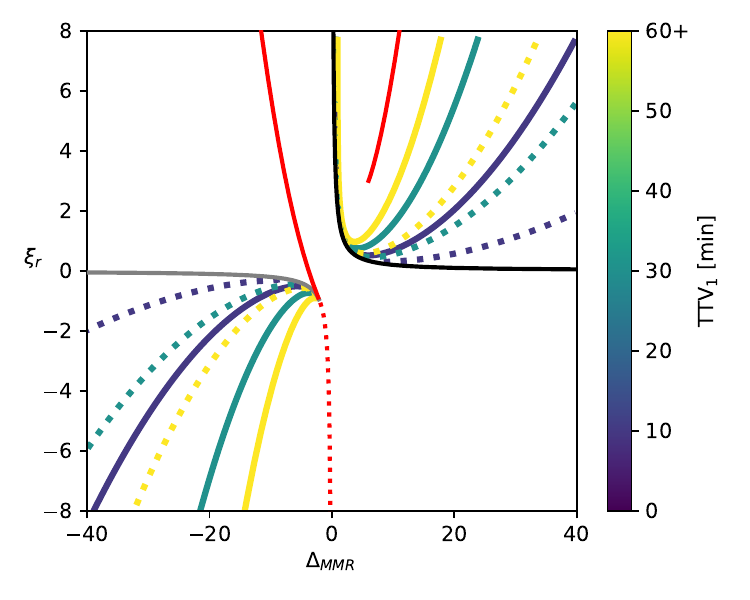}
\caption{TTV amplitudes as a function of position in the SFMR. Lines show the 10, 30 and 60 minute TTV amplitudes from the $\delta_{\lambda_1}$ component of the simplified analytical model (see Appendix \ref{ap:degen}) of a 10-day 5$m_\oplus$ planet around a sun-like star with a 15 day 5$m_\oplus$ perturber (solid line) or a 20d 1$m_\text{Jup}$ perturber (dotted line). Most of the resonant domain has an amplitude larger than 60 minutes.
}
\label{fig:TTVamp_bern} 
\end{center}
\end{figure}

 \begin{figure}
\begin{center}
\includegraphics[width=0.99\linewidth]{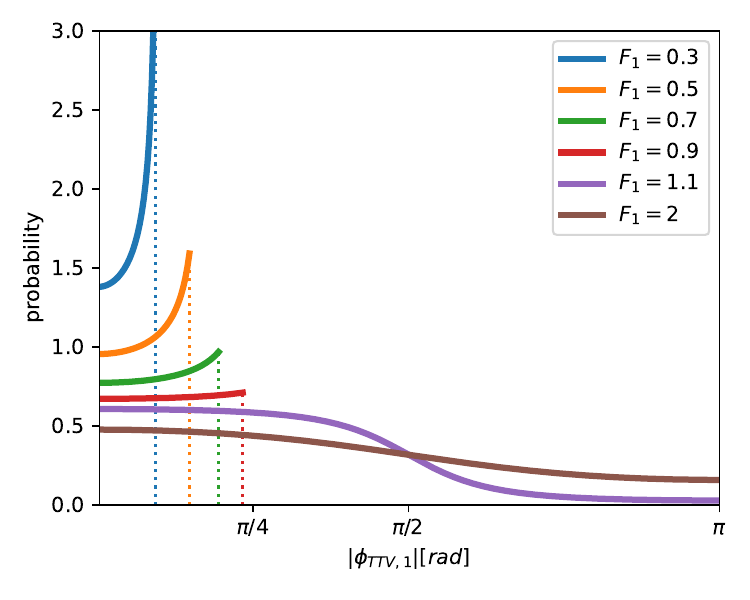}
\caption{Probability distribution of the TTV phase of the inner planet $\phi_{TTV,1}$ for different values of $F_1$.
}
\label{fig:phase_distrib} 
\end{center}
\end{figure}

The transit timing variations of a planet are, at first order in eccentricity, given by
\begin{equation}
\begin{aligned}
TTV_j = \frac{P_j}{2 \pi} \left( -\delta_{\lambda_j} +  \delta_{2 h_j} \right)\, ,
\end{aligned}
\label{eq:TTV}
\end{equation}
where $h_j =e_j \sin \varpi_j$, and $\delta_x$ are small variations of the quantity $x$, for which the constant component and the one evolving linearly with time are removed 
\citep{Nesvorny2009,NeVo2016}. Indeed, only non-linear temporal variations of $\lambda_j$ and  $e_j \sin \varpi_j$ induce TTVs, as linear evolution can be absorbed in a redefinition of the period. Examples of TTVs are shown on the bottom panel of Fig.~\ref{fig:elements}.

The TTV characterization of a pair of planets near a first-order MMR can be subject to degeneracies. Firstly, the secular eccentricity, $\es$, is constant in our model, and therefore does not impact the TTVs at first order. This is illustrated by having different shifts in the $(e_j \cos \varpi_j$, $e_j \sin \varpi_j)$ plane of the grey examples in the top panels, while having no substantive effect on the TTVs. In the case of Kepler-23, $\es$ evolves slightly over the 10 year integration, but in a manner that the evolution of $e_1 \sin \varpi_1$ is nearly linear. The upshot is that $\es$ is often poorly constrained \citep{HaLi2016,Leleu2023}. Secondly, the mass-eccentricity degeneracy \citep{Boue2012,Lithwick2012} is a degeneracy between the sum of the masses $M$ of the two planets and the resonant eccentricity $\er$.

To understand the $M$-$\er$ degeneracy in the ($\Delta_{\rm MMR}$,$\er$) plane, we adapted the analytical TTV model of \cite{NeVo2016}; see appendix \ref{ap:degen}. As described in \S \ref{sec:examples}, inside the resonance, $\delta_{\lambda_j}$ and $\delta_{2 h_j}$ evolve at different frequencies, while outside of the resonance, they both evolve with a frequency close to $\overline{f}_{super}$. In Fig. \ref{fig:SFM_degen}, the background color shows the area of the SFMR in which the $\delta_{\lambda_j}$ signal is degenerate (blue is subject to degeneracy), in the sense that estimating the amplitude, period and phase of the TTV signal with realistic precision does not enable one to position the pair in the SFMR, nor estimate the planets' masses. These degeneracies add to the (im-)precision of the determination of the observables: Fig.~\ref{fig:TTVamp_bern} shows that TTV amplitudes are typically smaller in that area, vanishing as we move away from the formally resonant domain.

The degeneracy is resolved inside the resonance due to the dependency of the resonant frequency on the planetary masses. Assuming a sinusoidal signal for both $\delta_{\lambda_j}$ and $\delta_{2 h_j}$, the cyan dashed curve shows the places where the sum of the TTV semi-amplitudes of $\delta_{\lambda_j}$ for the two planets $A_{TTV}=    (A_{TTV,1} + A_{TTV,2} ) $ is constant. Interestingly, we show that this curve depends on $\Delta_{\rm MMR}$ and $\er$ only. Measuring $A_{TTV}$ therefore provides a direct insight into the position of the pair of planets in the SFMR, drawing curves in the ($\Delta_{\rm MMR}$, $\er$) plane independently of the planetary masses (cyan dotted curves in Fig.~\ref{fig:SFM_degen}).

Outside of resonance, the relative amplitude of the terms is given by, for example for the inner planet: 
\begin{equation}
\begin{aligned}
F_1=\frac{\delta_{\lambda_1}}{\delta_{2h_1}} & \approx \frac{\er}{\Delta_{\rm MMR}} \sqrt{\frac{K_p}{2p}}M_p^{-1/3}\, .
 \end{aligned}
\label{eq:F1}
\end{equation}
$F_1$ therefore depends on $p$ and the planetary masses. 
Systems with relatively large $|\er|$ are therefore dominated by $\delta_{\lambda_1}$ and their posteriors are expected to follow the iso-amplitude curves as shown by the example of Fig. \ref{fig:SFM_degen}, samples with larger $|\er|$ having lower $M$.

$F_1$ also dictates the phase of the TTV signal $\phi_{TTV,1}$, defined as the difference between the phase at which the sinusoidal approximation of the TTV of the inner planet reaches zero while decreasing, and the phase at which the line of conjunction crosses the line of sight \citep[see appendix \ref{ap:phases},][]{Lithwick2012}. For $F_1=0$, the phase is 0, while it is bounded by a value below $\pi/4$ for $F_1\leq1$; see Fig. \ref{fig:phase_distrib}. The phase can take any value if $F_1>1$. For an individual system, therefore, the information gained from the phase is limited, although it was used to show that most pairs were not at the fixed points \citep{Lithwick2012,Choksi2023}. However, we can use the distribution of phases at a population level and see what distribution of $\xi_r$ (through the distribution of $F_1$), is required to reproduce the observed distribution of $\phi_{TTV,1}$, see \S \ref{sec:degen}.

To lift the $M$-$\xi_r$ degeneracies outside of resonance, additional constraints can be gained:
longer TTV baseline can help distinguish the resonant frequency from $\overline{f}_{super}$ \citep[][]{NeVo2016,Almenara2025}, while improved precision can also help in observing the chopping effect, or additional harmonics of the libration, especially closer to the resonant part of the phase space \citep{NeVo2016,Deck2015}.
Alternatively, measuring the mass of the planet using radial velocities will lift the  mass-eccentricity degeneracy.

Figure \ref{fig:elements} illustrates how $\es$ and $\er$ contribute to the full eccentricity vector through $e_{s,j}$ and $e_{r,j}$. Since $\es$ is often poorly constrained by TTVs, its posterior can be strongly influenced by the eccentricity prior and $\er$. Independent priors on $\er$ and $\es$ can alleviate this issue. Nonetheless, recent results showed that for the 2:1 MMR, $\es$ can influence the TTVs over the resonant timescale \citep[second order terms in eccentricities, see][]{Almenara2026}, while high precision might help to constrain the full eccentricity vector through the transit shape. Eventually, long baselines of tens of years will enable the observation of the non-linear evolution of $\vec{e}_{s,j}$ which will add a long period to the TTV signal.

To check whether the quantities potentially subject to degeneracies are well constrained by the data, one can perform analyses using different, but realistic, sets of priors and comparing the posteriors \citep{Jontof2016,HaLi2017,Leleu2023}. 
In the observational database, a flag indicates if the author tested the robustness of their analysis by trying different priors, and a second flag indicates if the mass published is robust to this change. By exploring the analyses that performed these tests in the database, we found that having robust mass determination typically implies having a robust estimation of $\er$, but the same is not true for $\es$, see Fig.~\ref{fig:obs_rob}.

\section{The resonant state of the observed population}
\label{sec:obs}
\subsection{Analysis selection}
\label{sec:selection}


We select pairs of planets that pass the criteria discussed in \S \ref{sec:choice}, \S \ref{sec:validity}, and with an outer orbital period of less than 70 days. 
The observation database can contain several analyses for a given system, which were provided by one or several teams. To choose a single analysis per system, we prioritize as follows: analyses with robustness checks are chosen over other analyses, then, photo-dynamical analyses are chosen over analyses of pre-extracted transit timings. If this filtering yields several analyses from a given author, the default flag is used. This yields a single analysis per system with the current database.

There are evidence of MMR-dependent trends in the Kepler dataset \citep[e.g.][]{Steffen2015}; however the number of pairs near each MMR, except the 3:2, is not large enough to study them individually. In this study, we therefore analyze the resonant state of all pairs, regardless of the particular first-order MMR to which they are associated. 


\subsection{Distribution in the SFMR}

We discuss the distribution of the observed pairs in the SFMR, shown in Fig. \ref{fig:SFMobs}. 
The most visible trend is that the robustly characterized pairs cluster near the primary fixed point curve.
Pairs clustered near the primary fixed points outside of the resonance-trapped domain are discussed in \S \ref{sec:Pmin}.

Planet pairs in, and near, the resonance-trapped domain should be robustly characterized\footnote{Except
if the pair is lying so close to the fixed point that the TTV amplitude at the resonant period cannot be estimated with the available precision, or if the large TTVs of small resonant planets prevent their detection in transit surveys \citep{Garcia2011,Leleu2021}.}, as they are not subject to the $M-\er$ degeneracy (see \S\ref{sec:degen}). We discuss their distribution in \S \ref{sec:reso}.

Most pairs in the upper or lower circulation area are affected by the $M-\er$ degeneracy. To study them, we therefore have to consider observables that alleviate it. In \S\ref{sec:upperDer} we use the TTV phase distribution at the population level to estimate the underlying distribution of $\Delta \er$ in the upper circulation area. In \S \ref{sec:fracextint}, we use mass-radius relationships to study the distance to first order MMRs for pairs that are too far for robust TTV characterization.

\subsubsection{Small $|\er|$ and dependency on the inner period}
\label{sec:Pmin}

\begin{figure}
\begin{center}
\includegraphics[width=0.99\linewidth]{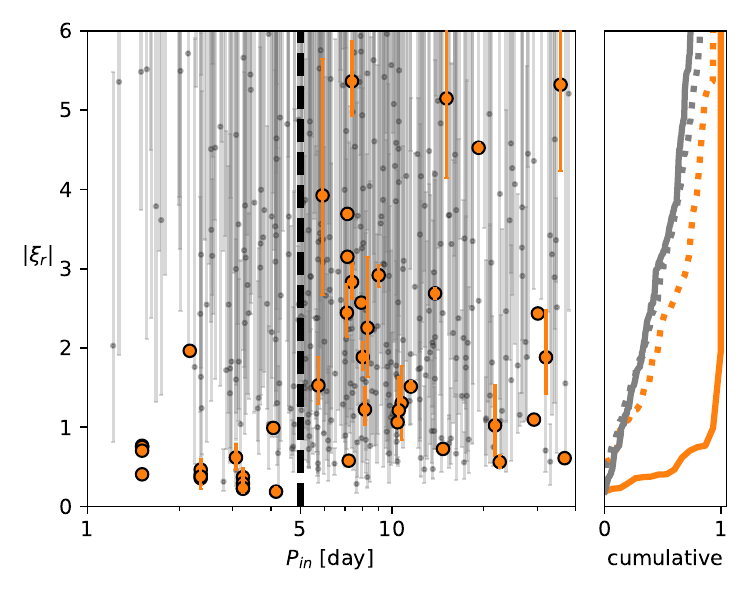}\\
\caption{$|\er|$ as a function of $P_{\rm in}$ with the same data as Fig.~\ref{fig:Pve1}. The vertical dashed line is at 5 days, which has been considered a limit at which tides play a major role in the system evolution \citep{Delisle2014letter,HaLi2017}. The cumulative curve shows in solid (resp. dotted) systems whose inner planet is below (resp. above) 5 days. Orange shows robustly characterized pairs, for which the difference between below and above $P_{\rm in}=5$d is significant (Kolmogorov-Smirnov statistic of $0.80$ with a p-value of $3.3\times10^{-7}$), while grey is the whole dataset (p-value of $0.15$). 
}
\label{fig:obs_Pinner} 
\end{center}
\end{figure}

The observed population shows an overabundance of planets near the fixed points at $\Delta_{\rm MMR}> 0$ and $\er \sim 0$ (Fig.~\ref{fig:SFMobs}). In Fig.~\ref{fig:obs_Pinner} we show $|\er|$ vs $P_{\rm in}$, which is either the period of the inner planet of the considered pair or the innermost planet if the pair is part of a known resonant chain. The result of the Kolmogorov-Smirnov (KS) test is strongest, both in terms of distance between the cumulative distributions (statistic), and the significance of the result (p-value) when placing the threshold between 4.2 and 5.7 days. A strong dependency of the eccentricity as function of the period of the inner planet is expected from planetary tides \citep[e.g.][]{Delisle2012}. This dichotomy with respect to a threshold at $\sim\!5$ days is therefore strong evidence for the effect of planetary tides on the close-in population, consistent with the distribution of period ratios near commensurability \citep{Delisle2014letter}, and the distribution of another component of the eccentricity vector \citep{HaLi2017}. We discard pairs with $P_{\rm in}<5$ days in the following sections.

\subsubsection{Distribution in the resonance-trapped domain}
\label{sec:reso}
\begin{figure}[h!]
\begin{center}
\includegraphics[width=0.95\linewidth]{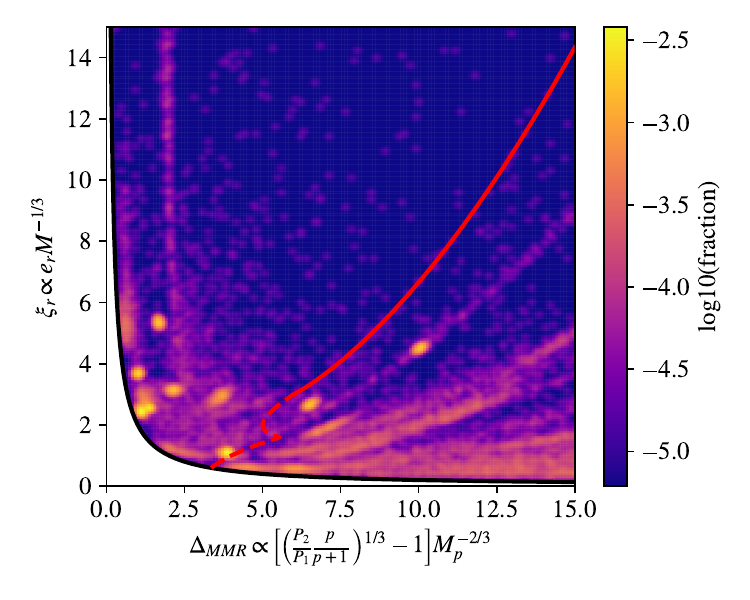}\\
\caption{Distribution of the posteriors in the resonant domain for $P_{\rm in}\!>\!5$ days. The color indicates the log of the fraction of samples in a given pixel. A 2D Gaussian filter of 1 pixel of standard deviation is applied to increase the visibility of well constrained pairs. 17 pairs have samples inside the formally resonant domain, in red in Fig \ref{fig:SFM_anarea} (a) and (b). 
}
\label{fig:reso} 
\end{center}
\end{figure}

Samples in the resonant domain or near its `entrance' with $P_2/P_1>(p+1)/p$ with an inner planet at more than $5$ days are shown in Fig. \ref{fig:reso}. Pairs inside the resonance or just outside with $\er \lesssim 4$ are less subject to the $\er-M$ degeneracy (see Fig. \ref{fig:SFM_degen}); all samples are therefore represented in Fig. \ref{fig:reso}.

Pairs inside the resonance-trapped domain ave moderate eccentricities 
$\smash{\er =3.66_{-2.54}^{+3.95}}$ (robust: $\smash{3.19_{-2.09}^{+2.65}}$), which is similar to the value at which the separatrix appears for $\delta=0$ ($\er=3$ Fig. \ref{fig:SFM_anarea} a). This correspond to a distribution of (non rescaled) resonant eccentricities of   $\smash{e_{r,j}=0.024_{-0.020}^{+0.037}}$ (robust: $\smash{0.020_{-0.019}^{+0.047}}$).
The pairs are also relatively close to the primary equilibria, with $\smash{\Delta \er =0.18_{-0.12}^{+0.27}}$ (robust: $\smash{0.11_{-0.06}^{+0.34}}$), while the separatrix is located at $\Delta \er=1$ for $\delta=0$. 

\subsubsection{Distribution of eccentricites for $P_2/P_1>(p\!+\!1)/p$}
\label{sec:upperDer}

\begin{figure}
\begin{center}
\includegraphics[width=0.99\linewidth]{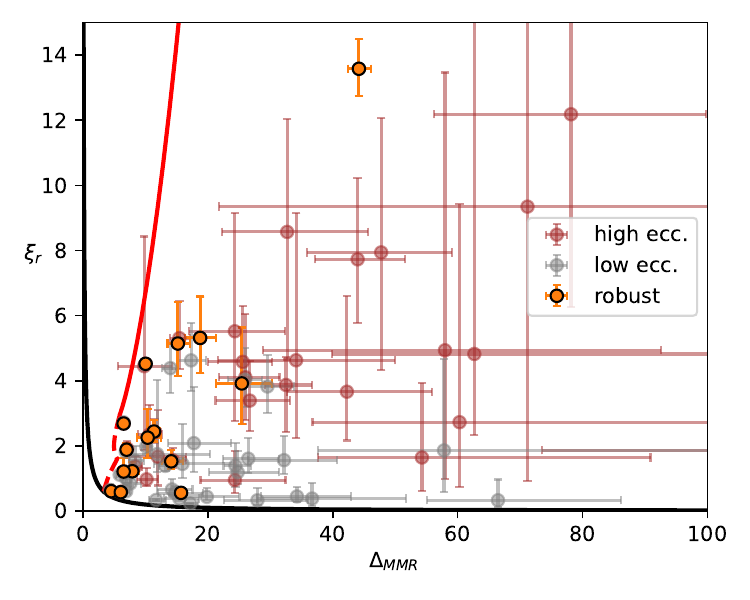}
\caption{Representation in the SFMR of non-resonant pairs with $P_{\rm in}\!>\!5$ days and $P_2/P_1>(p\!+\!1)/p$. The brown and grey points are from analyses that used uniform and log-uniform eccentricity priors, respectively \citep{HaLi2017,Leleu2023}. The orange points are the robustly characterized pairs. }
\label{fig:SFM_upper} 
\end{center}
\end{figure}

\begin{figure}
\begin{center}
\includegraphics[width=0.99\linewidth]{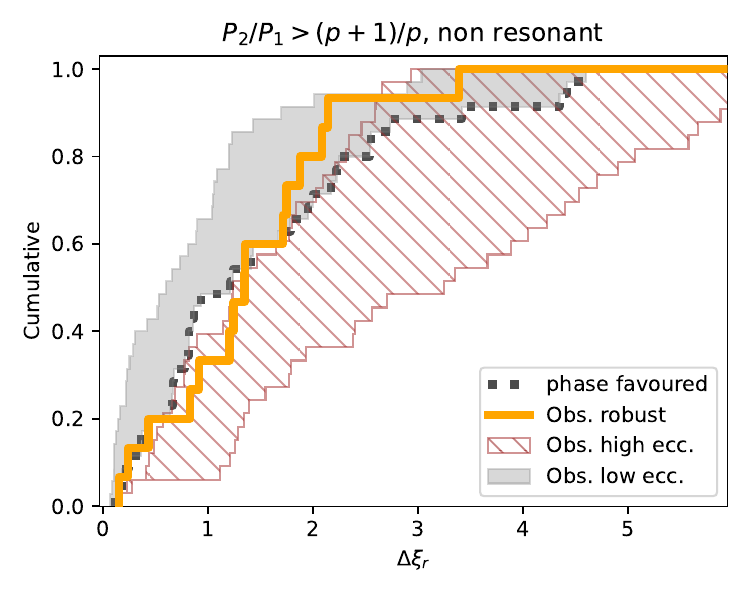}
\caption{Distribution of $\Delta \er$ for the pairs shown in Fig. \ref{fig:SFM_upper}, using the same color code.}
\label{fig:Der_upper} 
\end{center}
\end{figure}

Here we consider the `excitation' of the rescaled resonant eccentricity $\Delta \er$ outside of the resonance-trapped domain. We focus on the more populated side $P_2/P_1\!>\!(p\!+\!1)/p$. In Fig. \ref{fig:SFM_upper} we show 35 pairs that were analyzed using different mass and eccentricity priors (uniform eccentricity in brown and log-uniform in grey), while the robustly characterized pairs from the whole database are shown in orange. The robustly characterized pairs are closer to the separatrix, while the grey and brown posteriors populate a larger range of $\Delta_{\rm MMR}$. In Fig. \ref{fig:Der_upper} we show the cumulative distribution of $\Delta \er$($0.16-0.84$ quantiles) for the same populations.
The difference between the brown and grey populations highlights how this area of the SFMR is strongly affected by the $M-\er$ degeneracy. 

As described in \S\ref{sec:degen}, the TTV phases, $\phi_{TTV,j}$, can be used at the population level to constrain the distribution of $\Delta \er$. In Appendix \ref{ap:phases}, we 
show that samples selected at the 0.84 quantile of $\Delta\er$ of the log-eccentricity posterior best match the observed phase distribution (black dashed line in Fig.~\ref{fig:Der_upper}). Interestingly, this distribution is relatively close to the distribution of the robust population (K-S statistic of $0.22$, p-value of $0.62$). This implies that the distribution of $\Delta \er$ for pairs further away from resonance is consistent with that of the robust sample (the orange population has $\smash{\Delta_{\rm MMR}=10.39_{-3.84}^{+7.73}}$, while the grey population has $\smash{\Delta_{\rm MMR}=14.8_{-7.9}^{+14.1}}$). For the robust sample, the rescaled resonant eccentricities are $\smash{\er = 2.26_{-1.50}^{+2.74}}$, corresponding to a distribution of (non-rescaled) resonant eccentricities of $\smash{e_{r,j}=0.012_{-0.008}^{+0.018}}$ for the planets in these pairs.

\subsection{Distribution of near-resonant pairs }
\label{sec:fracextint}

\begin{figure}[h!]
\begin{center}
\includegraphics[width=0.95\linewidth]{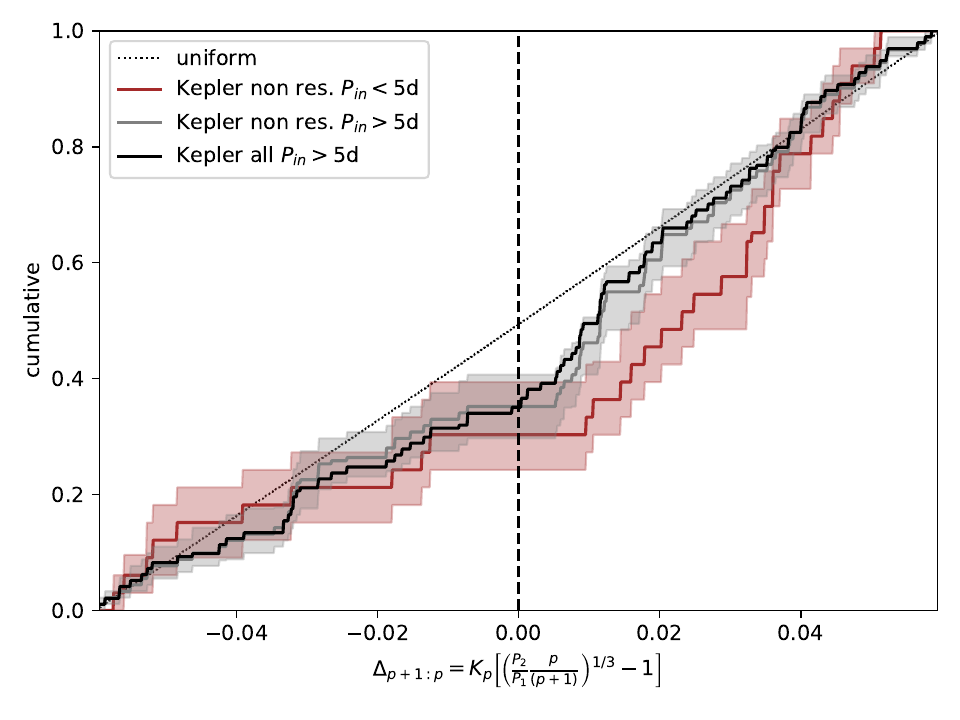}\\
\includegraphics[width=0.95\linewidth]{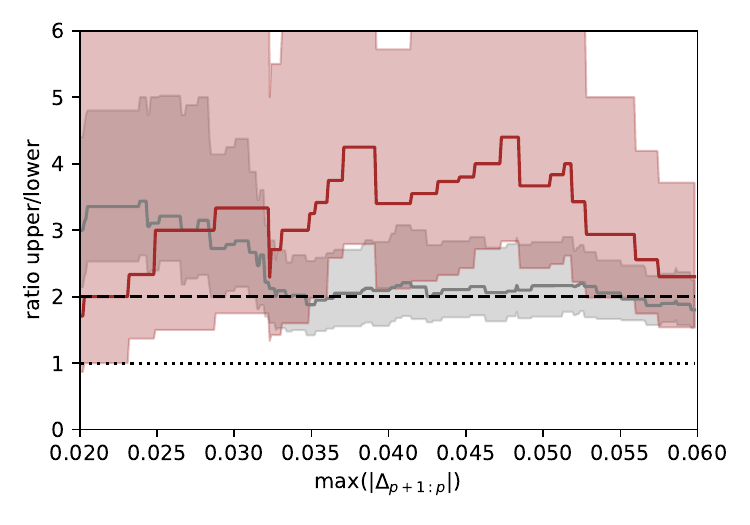}\\
\caption{Distribution of pairs in Kepler multi-planetary systems near the 3:2 and 2:1 MMRs.
$\Delta_{p+1:p}$ depends on the period ratios only (the constant $K_p$ depends on the chosen MMR, Table \ref{tab:Kp}). The shaded area represent the $1\sigma$ intervals estimated performing 1000 drawing with replacement of the considered population.
}
\label{fig:Dpp1p3221} 
\end{center}
\end{figure}

\begin{figure}[h!]
\begin{center}
\includegraphics[width=0.95\linewidth]{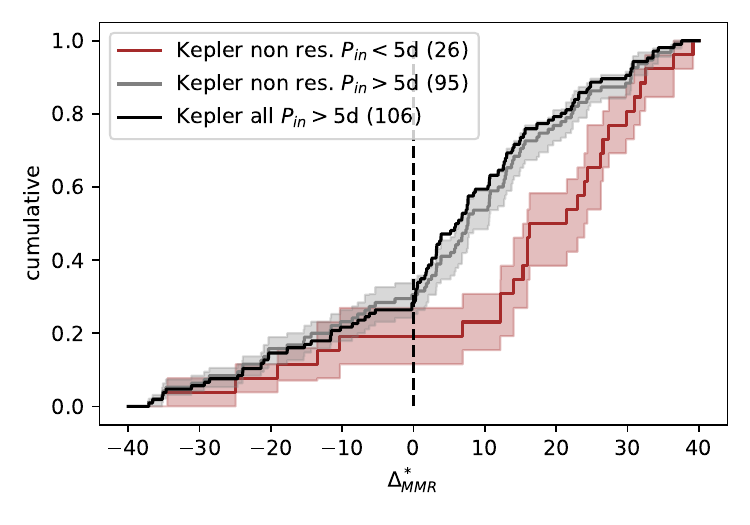}\\
\caption{Distribution of $\Delta_{\rm MMR}$ for all Kepler multis. The number of pairs for each distribution is given in parentheses. The $^*$ indicates that $\Delta_{\rm MMR}$ was estimated using a mass-radius relationship; see the text for more details.
}
\label{fig:DMMRall} 
\end{center}
\end{figure}

We now look at the overall distribution of pairs near first order resonances, as a function of distance to commensurability. In order to consider pairs further away from resonance than in the previous sections, and avoid TTV-related selection biases, we use the catalog of Kepler confirmed planets in the NASA exoplanet archive, and quantities that can be derived from that catalog. Figure \ref{fig:Dpp1p3221} shows the cumulative distribution for pairs near the 3:2 and 2:1 MMRs. These MMR are chosen because the relatively wide symmetrical domains around these MMRs make it possible to see how the near-resonant domain reconnects with the continuum : the limit of $|\Delta_{p+1:p}|$ of 0.06 is set by the width of the symmetric domain around the most compact of the considered MMRs (3:2 in this case, see Fig. \ref{fig:Dpp1p}). The ratio of the number of pairs closer to MMRs ($|\Delta_{p+1:p}|\!\lesssim\!0.03 $) over the number of pair further away ($ 0.03\!\lesssim\!|\Delta_{p+1:p}|\lesssim 0.06$) is $\smash{0.90_{-0.14}^{+0.17}}$, implying that near-resonant pairs are neither over- nor under-abundant as a whole
\footnote{The rest of the results of this section hold when restricting to pairs in the vicinity of the 3:2 MMR, although they represent an over density with respect to the continuum \citep[ratio of $\smash{1.49_{-0.31}^{+0.35}}$, see also][]{Steffen2015}. }. 
Splitting the pairs as a function of the period of their innermost planet, this ratio is $\smash{1.05_{-0.19}^{+0.23}}$ for the far-out ($P_{\rm in}\!>\!5$d - black) population and $\smash{0.57_{-0.18}^{+0.23}}$ for the close-in population ($P_{\rm in}\!<\!5$d - brown).
Both distributions significantly depart from the uniform one for $|\Delta_{p+1:p}|\!\lesssim \!  0.03$, with an under-abundance below commensurability, and an over-abundance above commensurability, as noted by \cite{Fabrycky2014}. To go further, we use the \texttt{resonantstate} database to identify pairs that are in the resonance-trapped domain, and remove them from the sample.

The fraction of non-resonant pairs above and below commensurability as a function of maximum distance, $\Delta_{p+1:p}$, is shown in the bottom panel of Fig. \ref{fig:Dpp1p3221}. This fraction is always above one, and broadly consistent between the close-in ($P_{\rm in}\!<\!5$d - brown) and far-out ($P_{\rm in}\!>\!5$d - grey) populations. However, a significant fraction of the close-in pairs above commensurability seem to have a larger departure from $\Delta_{p+1:p}=0$. 

To investigate this further, we leverage the rescaling of the SFMR by estimating $\Delta_{\rm MMR}$ of all consecutive pairs (near any $p+1:p$ MMR) of confirmed Kepler multi-planetary systems such that $P_2\!<\!70$d, using the mass-radius relationship from \cite{Parc2024} and stellar parameters from \cite{Berger2020}. The resulting distributions are shown in Fig. \ref{fig:DMMRall}. Using a Kolmogorov-Smirnov test, we show that the close-in and farther out populations statistically differ with a maximum distance of $0.369$ located at $\Delta_{\rm MMR}=11.84$, with a p-value of $0.005$, consistent with a shift towards larger distance to commensurability for close-in pairs with $\smash{P_2/P_1>(p+1)/p}$ \citep{Delisle2014letter}.

\section{Discussion and Conclusion}
\label{sec:conclusion}
We presented the \texttt{resonantstate} observation database, which compiles posterior samples for 1714 planets in 675 systems, and projected pairs near first-order mean motion resonances (MMRs) onto the Second Fundamental Model of Resonance. This describes their resonant state independently of the planetary masses and of the specific commensurability, and identifies where the mass-eccentricity degeneracy limits TTV characterization. As expected, most of the robustly-characterized pairs (pairs whose resonant state does not depend strongly on the choice of prior, hence not subject to the mass-eccentricity degeneracy) are either in the resonance-trapped domain, or its immediate vicinity. The neighborhood of first-order MMRs can still be analyzed at a population level, using either the distribution of observed TTV phases, or mass-radius relationships. Since $\Delta_{\rm MMR}$ depends only on instantaneous orbital periods and planetary masses, RV follow-up of transiting near-resonant pairs would enable a better estimation of these quantities in the future.  

Regarding far-out pairs ($P_{\rm in}\!>\!5$d), those in the Resonance-trapped domain have moderate rescaled resonant eccentricities ($\er = 3.19_{-2.09}^{+2.65}$), comparable to the value at which the separatrix appears ($\er = 3$). Assuming no significant evolution after the disk dispersal, this value depends on the ratio between the eccentricity damping timescale and the migration timescale, which depends on the  aspect ratio of the disk \citep{TaTaWa2002,TaWa2004}, and the viscosity of the disk \citep{Pichierri2023,Pichierri2024d}. The pairs also tend to be close to the primary equilibria. Convergence toward the primary fixed points is achievable in proto-planetary disks in absence of strong turbulence \citep[e.g.][]{GoBa2023}, and in the absence of over-stable libration, which also depend on the disk parameters \citep{GoSc2014,Afkanpour2024}. Pairs outside of the resonance were found to also have non-negligible eccentricities \citep{Lithwick2012,HaLi2017,Choksi2023}, which we quantified to be similar to, although potentially slightly lower than, the resonance-trapped population ($\er = 2.26_{-1.50}^{+2.74}$). This is inconsistent with their formation in a strong eccentricity damping environment and could be consistent with the breaking of former chains, which we will discuss in the second paper in the series. Finally, we confirmed that only a few pairs were in the resonance-trapped domain, and that the asymmetry between pairs above and below exact commensurability comes from near-, but non-resonant pairs. Based on the departure from a uniform distribution, we defined this near-resonant domain as $|\Delta_{p+1:p}|<0.03$ or $|\Delta_{\rm MMR}|<40$. Within that domain, the ratio of pairs above vs below exact commensurability is $\smash{2.47^{+1.1}_{-0.6}}$.

Close-in systems ($P_{\rm in}\!<\!5$d) have even fewer resonance-trapped pairs, and none are found in the Kepler sample. The distribution of $\er$ is significantly lower than in the far-out population ($\smash{\er=0.39_{-0.15}^{+0.31}}$), in agreement with previous work \citep{HaLi2017}. Near-resonant pairs ($|\Delta_{p+1:p}|<0.03$) appear to be somewhat of an under-density compared to the continuum. This could be due to these pairs completing a larger number of orbits over a given system age, which could make them more subject to long-term instabilities. However, there is an additional significant difference between the close-in and far-out population: the close-in population above commensurability is significantly shifted towards a larger departure from commensurability \citep[see \S \ref{sec:fracextint}, and][]{Delisle2014letter}. However, the ratio of non-resonant pairs above/below commensurability (and the over-density of pairs above, but outside, the resonances) is comparable between both populations.

Since this over-density is present in both the close-in and far-out populations, and the far-out population is retained non-zero resonant eccentricities, it is unlikely that this feature as a whole can be explained by tides \citep{Lee2013,Delisle2014}. We will discuss the origin of the asymmetry in the second paper of the series.    
The similarities and discrepancies between the close-in and far-out populations are however consistent with the architecture of the far-out population being the starting point of a long-term tidal evolution that would result in the close-in population : 
both the damping of the resonant eccentricity and an increased distance from exact commensurability of about a percent in period ratio, hence in $\Delta_{p+1:p}$, is can happen over Gyrs timescales for $P_{\rm in}<5$d \citep{LiWu2012,Delisle2012}.
Having many more pairs above commensurability for the close-in population could have implied that a large number of formally resonant pairs smoothly evolved by tides following the primary fixed point and remained just outside of the resonance-trapped domain. Since the ratio of pairs above/below commensurability is not much larger for the close-in pairs, it implies that the mechanisms that generated the over-abundance of pairs above commensurability with non-zero eccentricities mostly took place before the systems had time to evolve through tides. The second paper in the series will compare the resonant state of observed pairs to the outcome of simulations to constrain planetary formation pathways. 

Some of the trends discussed in this paper rely on 10 to 20 robustly characterized pairs. To strengthen the results, enable studies of sub-populations (per type of planet, per resonances, etc.), as well as broaden the scope to system further away from resonance, we welcome\footnote{Please contact adrien.leleu@unige.ch.} new additions of posteriors of dynamical analysis (i.e. orbital elements at epoch) of multi-planetary systems regardless of their proximity to resonance.

\bibliographystyle{aa}
\bibliography{biblio}

\appendix
\nolinenumbers

\begin{acknowledgements}
AL, YE, JC, JK and MR acknowledge support of the Swiss National Science Foundation under grant number TMSGI2\_211697. This work has been carried out within the framework of the NCCR PlanetS supported by the Swiss National Science Foundation under grants 51NF40\_182901 and 51NF40\_205606. SH acknowledges support by the Natural Sciences and Engineering Research Council of Canada (NSERC), funding references CITA 490888-16 and RGPIN-2020-03885. MG acknowledges support from the ERC project No. 101019380 "HolyEarth.". National Aeronautics and Space Administration Exoplanet Research Program (NASA XRP) grant 80NSSC21K1111. This work was performed in part by the Virtual Planetary Laboratory Team, which is a member of the NASA Nexus for Exoplanet System Science, and funded via the NASA Astrobiology Program ICAR Grant 80NSSC23K1398. DR is supported by the National Science Foundation under Award No. 2143195. JK acknowledges support from the Spanish Research Agency of the Ministry of Science, Innovation and Universities (AEI-MICIU) under grant `Contribution of the IAC to the PLATO Space Mission' with reference PID2023-149439NB-C41. KM was supported by JSPS KAKENHI grant No 25K07387. GP acknowledges support from the European Union (ERC Starting Grant DiscEvol, project number 101039651) and from Fondazione Cariplo, grant No. 2022-1217. Views and opinions expressed are, however, those of the author(s) only and do not necessarily reflect those of the European Union or the European Research Council. Neither the European Union nor the granting authority can be held responsible for them. LBo acknowledges support from CHEOPS ASI-INAF agreement n. 2019-29-HH.0. ACMC acknowledges support from FCT, Portugal, through the CFisUC project UID/04564/2025 (with DOI identifier 10.54499/UID/04564/2025). FD acknowledges support by NSF CAREER Award 2441950. IGJ acknowledges support of National Science and Technology Council in Taiwan under grant NSTC 114-2112-M-007-029. ACP has been supported by the French government, through the $ \mathrm{UCA^{J.E.D.I.}} $ Investments in the Future project managed by the National Research Agency (ANR) with the reference number ANR-15-IDEX-01 MW acknowledges support of the National Natural Science Foundation of China (grant No.124B2058). TGW acknowledges support from the University of Warwick and UKSA. DAY is supported by a Juan Carlos Torres Postdoctoral Fellowship at the Massachusetts Institute of Technology. This work was granted access to the HPC resources of MesoPSL financed by the Region Île-de-France and the project Equip@Meso (reference ANR-10-EQPX-29-01) of the  programme Investissements d’Avenir supervised by the Agence Nationale pour la Recherche. LCY acknowledges support of National Science and Technology Council in Taiwan under grant NSTC 114-2115-M-007-007.  EA acknowledges support from NASA XRP grant 80NSSC21K1111 and NASA ICAR Grant 80NSSC23K1398.
\end{acknowledgements}

\section{Presentation of the databases}
\label{ap:DB}
\subsection{Overview}

{The databases are comprised of analysis and simulations files, which contain samples of the MCMC analysis of an observed system, and the time evolution of a simulation, respectively. These files are then referenced in metadata tables that help the user to select the analysis they are interested in, as well as information on how a given analysis was performed, and who to credit for a given analysis.} 

{The osculating astrocentric orbital elements used are: Mean longitude $\lambda_j$ in degrees, Period $P_j$ in days, $k_j = e_j \cos(\varpi_j)$ and  $h_j = e_j \sin(\varpi_j)$ where $e_j$ is the eccentricity and $\varpi_j$ the longitude of the periapsis, Inclination $I_j$ in degrees, Longitude of the ascending node $\Omega_j$  in degrees.}

\subsection{Structure}

{The bulk of the data is stored in \textit{analysis} files for samples of MCMC analysis of observational data, and  \textit{simulation} files for the temporal evolution of a simulation. Each of these files is referred to with a unique identifier. For the observation database, the identifier is made up of the name of the star and an analysis integer. For the simulation database, the identifier is made up of the simulation type, a run integer and a simulation integer.} 

{Analysis files contain MCMC samples, each sample having a unique integer identifier given in Column 0. Simulation files contain the temporal evolution of a simulation, each line corresponding to a time stamp which is given in Column 0. For a system with $N$ planets, the columns $1$ to $8N+2$ are the same for both file types:}

\begin{itemize}
\item Columns $8j+1$: Mean longitude $\lambda_j$ of planet j in degrees, 
\item Columns $8j+2$: Period of planet j in days, 
\item Columns $8j+3$: $k_j = e_j \cos(\varpi_j)$,
\item Columns $8j+4$: $h_j = e_j \sin(\varpi_j)$ ,
\item Columns $8j+5$: Inclination $I_j$ of planet j in degrees, 
\item Columns $8j+6$: Longitude of the ascending node $\Omega_j$ of planet j in degrees, 
\item Columns $8j+7$: $m_j/M_*$, where $m_j$ is the mass of planet j and $M_*$ is the stellar mass, 
\item Columns $8j+8$: $R_j/R_*$ Radii of the planets divided by stellar radius,
\item Column  $8N+1$: Mass of the star in Solar masses, 
\item Column  $8N+2$: Radius of the star in Solar radii, 
\end{itemize}
{where $e_j$ is the eccentricity and $\varpi_j$ the longitude of the periapsis of planet $j$. Optional additional columns are defined either in the `readme' file or the `additional info' field of the metadata table. }

\subsubsection{Observations metadata}
\label{ap:db_obs}
{The observation metadata table\footnote{\url{https://dace.unige.ch/resonantStateObservationsCatalog/}} contains a line per planet per analysis, implying that a given planet can appear on several lines if the system was analysed more than once. Each line contains the star name, the planet name, the link to the \textit{analysis} file, the publication(s) to cite when using the data, the median of the planetary and stellar parameters for the given planet and the given analysis, the reference epoch of the orbital elements, as well as additional information such as the priors used for the eccentricities and masses, the method used (e.g. `TTVs' or `photo-dynamical' analysis), the instruments, etc.}

\subsubsection{Simulations metadata}

{For the simulation database, the metadata are broken down into two layers: a run metadata table\footnote{\url{https://dace.unige.ch/resonantStateSimulationsCatalog/}} , and simulation metadata tables. 
This structure was adopted because simulations are often performed in runs. A run is defined as a set of simulations which have identical physical assumptions and models, with different initial conditions (e.g. randomized, grid-based, etc.).} 

{Each run correspond to a line in the run metadata table. The columns contain the simulation type (e.g. hydrodynamical, population synthesis, protoplanetary disk, tidal evolution, etc.), an integer identifier for each run of a given simulation type, the link to the corresponding simulation metadata table, the physics implemented, the publication(s) to cite when using the data, the  name of the author, etc. 
The simulation metadata table has a structure similar to the observations metadata table: a line per planet, containing the planetary and stellar parameter at the end of the simulation, a unique simulation integer identifier, the link to the associated simulation file, as well as additional information on how the simulation was performed.}

\section{Derivation of the SFMR}
\label{ap:SFM}
{The SFMR is an integrable Hamiltonian system that approximately describes the dynamics of pairs of planets near first-order mean-motion resonances. Through a series of canonical transformations and rescalings, it provides a universal description of the dynamics that is independent of both the particular $p+1\!:\!p$ resonance under consideration and planets’ masses.}

{We consider two planets of masses $m_1$ and $m_2$ orbiting a star of mass $m_\star$, close to a first-order MMR $p+1:p$. We denote $\iota=\sqrt{-1}$ and $\,^t\cdot$ the transpose operator.}

\subsection{Hamiltonian of the problem}\label{ap:Hamiltonian}

{Starting from the canonical heliocentric coordinates of \cite{LaRo1995}, we use the Poincar\'e canonical coordinates: 
\begin{equation}\label{eq:poincare}
\begin{aligned}
\Lambda_j & =\beta_j\sqrt{\mu_ja_j} \,  , & \lambda_j  \\
D_j & =\Lambda_j\left(1-\left(1-e_j^2\right)^{1/2}\right)\, , & -\varpi_j
\end{aligned}
\end{equation}
where the momenta $\Lambda_j$ and $D_j$ are canonically conjugate to coordinates $\lambda_j$ and $-\varpi_j$, respectively. The Hamiltonian is
\begin{equation}
	H=-\sum_{1\leq j\leq2}\frac{\beta_j^3\mu_j^2}{2\Lambda_j^2}+H_P(\Lambda_j,D_j;\lambda_j,-\varpi_j),
\end{equation}
where $\mu_j=\G\left(m_\star+m_j\right)$ and $\beta_j=m_\star m_j/\left(m_\star+m_j\right)$. }
{At first order in eccentricity and averaging over fast angles, noting , $H_P$ becomes \citep[e.g.][]{DePaHo2013}:}
\begin{equation}\label{Pert_Poincare}
	\begin{split}
		&H_P=-\frac{Gm_1m_2}{a_2}\left(f_0+f_1\sqrt{\frac{2D_1}{\Lambda_1}}\cos[p\lambda_1-\pp\lambda_2+\varpi_1]\right.\!\!\!\!\!\!\!\!\!\!\!\!\!\!\!\\
		&\left.+f_2\sqrt{\frac{2D_2}{\Lambda_2}}\cos[p\lambda_1-\pp\lambda_2+\varpi_2]\right)+\OO\left(e_j^2\right).
	\end{split}
\end{equation}
where $f_0$, $f_1$ and $f_2$ are functions of $\alpha_{12}=a_1/a_2$ estimated at the exact commensurability of a given resonance $p$. $f_0$ is dropped as it is in factor of a term considered constant.
\begin{equation}\label{eq:f1f2}
	\begin{split}
		f_1 &= - \left(p+1+ \frac{\alpha_{12}}{2} \frac{d}{d\alpha_{12}} \right) b^{(p+1)}_{1/2} \left(\alpha_{12}\right),\\
        f_2 &=   \left(p+\frac{1}{2}+\frac{\alpha_{12}}{2} \frac{d}{d\alpha_{12}} \right) b^{(p)}_{1/2} \left(\alpha_{12}\right)-\delta_{1,p}\alpha_{12}^{-1/2},
	\end{split}
\end{equation}
where $\delta_{x,y}$ is $1$ if $x=y$ and $0$ otherwise. The \smash{$b_s^{(k)}$} are the Laplace coefficients (Laplace 1798, Poincaré 1905) defined as \citep[][pages 496-505]{BrouwerClemence1961}
\begin{equation}
    \begin{split}
        b_s^{(k)}(\alpha)&=\frac{2}{\pi}\int_0^\pi\left(1-2\alpha\cos\theta+\alpha^2\right)^{-s}\cos k\theta\,d\theta,\\
        \frac{db_{s}^{(k)}}{d\alpha}&=s\left(b_{s+1}^{(k-1)}(\alpha)-2\alpha b_{s+1}^{(k)}(\alpha)+b_{s+1}^{(k+1)}(\alpha)\right).
    \end{split}
\end{equation}
\subsubsection{Reduction to two degrees of freedom}
The Hamiltonian is readily reduced to two degrees of freedom with the transformation \citep[\textit{e.g.}][]{Delisle2017}:
\begin{equation}
\begin{aligned}
&\Gamma  =(p+1) \Lambda_1 + p \Lambda_2 \, ,  & \, \varphi_1&=\lambda_1 - \lambda_2  \\
&G  = \Lambda_1 + \Lambda_2 - D_1 - D_2 \, ,  & \, \varphi_2&= - p \lambda_1 +(p+1) \lambda_2  \\
&D_1 \, ,& \, \sigma_1  &= - p \lambda_1 +(p+1) \lambda_2  -\varpi_1 \\
&D_2 \, ,& \, \sigma_2 &= - p \lambda_1 +(p+1) \lambda_2  -\varpi_2 \\
\end{aligned}
\end{equation}
The angles $\varphi_1$ and $\varphi_2$ do not appear in the Hamiltonian, which is reduced to the two degrees of freedom $\left(D_1,D_2;\sigma_1,\sigma_2\right)$. Both $\Gamma$ (combination of circular angular momenta) and $G$ (total angular momentum) are constant parameters.

\subsubsection{Expression of the Keplerian part}

The Keplerian part is expanded to second order in $\Delta\Lambda_j=\Lambda_j-\Lambda_{j,0}$. We obtain
\begin{equation}\label{eq:HamKep}
	H_K=\sum_{j\in\left\lbrace1,2\right\rbrace}\left(n_{j,0}\Delta\Lambda_j-\frac{3}{2}n_{j,0}\frac{\Delta\Lambda_j^2}{\Lambda_{j,0}}\right).
\end{equation}
We define the parameters $\Delta\Gamma$ and $\Delta G$ by
\begin{equation}\label{eq:DG}
	\begin{aligned}
		&\Delta\Gamma=\Gamma-\left(p+1\right)\Lambda_{1,0}-p\Lambda_{2,0},\\
		&\Delta G=G-\Lambda_{1,0}-\Lambda_{2,0},
	\end{aligned}
\end{equation}
Without loss of generality, we can choose the reference semi-major axis such that $\Delta\Gamma=0$. We obtain
\begin{equation}\label{final_Keplerian}
	\begin{aligned}
\Ham_K'=- \frac{3}{2}n_{1,0} p   \left(\Delta G+D_1+D_2 \right)^2 \left( \frac{p}{\Lambda_{1,0}}+\frac{p+1}{\Lambda_{2,0}} \right) \, .
	\end{aligned}
\end{equation}

\subsubsection{Expression of the perturbative part}
The system is assumed to remain close to the resonance $pn_1-\left(p+1\right)n_2=0$. Therefore, we define nominal mean motions $n_{j,0}$ such that $p n_{1,0}=\left(p+1\right)n_{2,0}$ and nominal semi-major axes $a_{j,0}$ and $\Lambda_{j,0}$ with
\begin{equation}
	\mu_j=n_{j,0}^2a_{j,0}^3\text{ and }\Lambda_{j,0}=\beta_j\sqrt{\mu_ja_{j,0}}.
\end{equation}
We expand the perturbative part of the Hamiltonian to order zero in $\Delta\Lambda_j=\Lambda_j-\Lambda_{j,0}$ and the Keplerian part to order two because both expansions generate a remainder of the same size. For the perturbation, an expansion to order zero means that $\Lambda_j$ is evaluated at $\Lambda_{j,0}$. Once evaluated at \smash{$a_1/a_2=\left(p/\pp\right)^{2/3}$}, the coefficients $f_j(a_1/a_2)$ become constants depending on $p$ and $f_0$ is removed from the Hamiltonian. We give $f_1$ and $f_2$ in Table \ref{tab:f_j} for $1\leq p\leq5$.
\begin{table*}
	\begin{center}
		\begin{tabular}{cccccc}
			\hline
			\hline
			$p$&$1$&$2$&$3$&$4$&$5$\\
			\hline
			$-f_1$&$1.1904936978$&$2.0252226899$&$2.8404318567$&$3.6496182441$&$4.4561427851$\\
			$f_2$&$0.4283898341$&$2.4840051833$&$3.2832567218$&$4.0837053718$&$4.8847062975$\\
			\hline
			\hline
		\end{tabular}
		\caption{Coefficients $f_j$ appearing in Eq. (\ref{Pert_Poincare}) and given in Eq. (\ref{eq:f1f2}) evaluated at $\alpha_{12}=\left(p/\pp\right)^{2/3}$.}\label{tab:f_j}
	\end{center}
\end{table*}
The perturbation now reads
\begin{equation}\label{eq:Hp_sigma_d}
	\Ham_P'\!=\!-\frac{m_1n_{2,0}\Lambda_{2,0}}{m_\star}\!\left(f_1\sqrt{\frac{2D_1}{\Lambda_{1,0}}}\cos\sigma_1\!+\!f_2\sqrt{\frac{2D_2}{\Lambda_{2,0}}}\cos\sigma_2\right)\!.
\end{equation}

\subsection{Reduction to one degree of freedom}\label{sec:1dof}

The Hamiltonian $\Ham'=\Ham_K'+\Ham_P'$, has two degrees of freedom. However, there exists another conserved quantity, in addition to $G$ and $\Gamma$, which allows one more degree of freedom to be lost by performing a rotation of the set of variables \citep{SeFe1984}. We define the Cartesian canonical coordinates:
\begin{equation}
		u_j=\sqrt{2D_j}\cos\sigma_j,\;\;v_j=\sqrt{2D_j}\sin\sigma_j.
        \label{eq:uv}
\end{equation}
In these coordinates, the Hamilton equations are $\dot{u}_j=-\partial\Ham'/\partial v_j$ and $\dot{v}_j=\partial\Ham'/\partial u_j$. The expression of $\Ham_K'$ is unchanged but $D_1+D_2=\frac{1}{2}\left(u_1^2+v_1^2+u_2^2+v_2^2\right)$. The perturbative part now reads:
\begin{equation}
	\Ham_P'\!=\!-\frac{m_1n_{2,0} \Lambda_{2,0}}{m_\star}\!\left(\frac{f_1}{\sqrt{\Lambda_{1,0}}}u_1\!+\!\frac{f_2}{\sqrt{\Lambda_{2,0}}}u_2\right)\!=\!\alpha_1u_1\!+\!\alpha_2u_2.
\end{equation}
Let $U,V\in\mathbb{C}$ and $\phi$ an angle defined as
\begin{equation}
	U=u_1+\iota u_2,\;\;\;V=v_1+\iota v_2,\;\;\;\phi=\arctan2(\alpha_2,\alpha_1).
\end{equation}
We now perform the canonical rotation \citep{SeFe1984,Henrard1986}:
\begin{equation}
	U=e^{\iota\phi}\tilde{U},\;\;\;V=e^{\iota\phi}\tilde{V},
    \label{eq:UV}
\end{equation}
where $\tilde{U}=x_1+\iota x_2$ and $\tilde{V}=y_1+\iota y_2$. The Keplerian part is still unchanged but $d_1+d_2=\frac{1}{2}\left(x_1^2+y_1^2+x_2^2+y_2^2\right)$. As for the perturbative part, using the identities $\cos(\phi)=\alpha_1/\sqrt{\alpha_1^2+\alpha_2^2}$ and $\sin(\phi)=\alpha_2/\sqrt{\alpha_1^2+\alpha_2^2}$, it now reads
\begin{equation}
	\Ham_P'=\frac{m_1n_{2,0}\Lambda_{2,0}}{m_\star}\sqrt{\frac{f_1^2}{\Lambda_{1,0}}+\frac{f_2^2}{\Lambda_{2,0}}}\,x_1.
\end{equation}
We go back to canonical polar coordinates by writing
\begin{equation}
	\begin{split}
		&x_1=\sqrt{2R'}\cos r,\;\;\;y_1=\sqrt{2R'}\sin r,\\
		&x_2=\sqrt{2S'}\cos s,\;\;\;y_2=\sqrt{2S'}\sin s.
	\end{split}
    \label{eq:xjyj}
\end{equation}
Once again, the Keplerian part is still given by Eq. (\ref{final_Keplerian}) in these coordinates but $D_1+D_2=R'+S'$. The perturbative part reads
\begin{equation}
	\Ham_P'=\frac{m_1n_{2,0}\Lambda_{2,0}}{m_\star}\sqrt{\frac{f_1^2}{\Lambda_{1,0}}+\frac{f_2^2}{\Lambda_{2,0}}}\sqrt{2R'}\cos r.
\end{equation}

The Hamiltonian does not depend on the angle $s$, which means that its conjugated action $S'$ is a first integral. 
The Hamiltonian is now reduced to the single degree of freedom $\left(R';r\right)$. We now rewrite the full Hamiltonian:
\begin{equation}\label{eq:secondFundamentalModel}
	\Ham'(R';r)=\alpha R'-\beta R'^2+\gamma\sqrt{2R'}\cos r,
\end{equation}
where
\begin{equation}
	\begin{split}
		&\alpha=-2 \beta \left(\Delta G+S' \right) \!
        ,\!\!\!\!\!\!\!\!\!\!\!\!\!\!\!\!\!\!\!\!\!\!\\
		&\beta=\frac{3}{2}n_{1,0}p\left(\frac{p}{\Lambda_{1,0}}\!+\!\frac{p+1}{\Lambda_{2,0}}\right),\\
		&\gamma=\frac{m_1n_{2,0}\Lambda_{2,0}}{m_\star}\sqrt{\frac{f_1^2}{\Lambda_{1,0}}+\frac{f_2^2}{\Lambda_{2,0}}}.
	\end{split}
\label{eq:abg}
\end{equation}

\subsection{Reduction to one parameter}\label{sec:1parameter}

Equation \ref{eq:secondFundamentalModel} is the SFMR of resonance of Jacques Henrard and Anne Lemaître. It depends on the three parameters $\alpha$, $\beta$ and $\gamma$. However, there exists a classical method to reduce the Hamiltonian (\ref{eq:secondFundamentalModel}) to one single parameter $\delta$. Following \cite{HenLe1983}, we define $\sigma=r,R=R'/K,\Ham=\Ham'/\omega$ and $S=S'/K$
and we choose $K$ and $\omega$ in such a way that\footnote{Putting a $2$ in front of $\sqrt{2R}\cos r$ or a $3\left(\delta+1\right)$ in front of $R$ is arbitrary. We made this choice to have  the same Hamiltonian as \cite{HenLe1983}.
} the coefficient in front of $R^2$ in $\Ham$ be $-1$ and that in front of \smash{$\sqrt{2R}\cos r$} be $2$. This gives 
\begin{equation}
K=\left(\gamma/2\beta\right)^{2/3} \, \text{and}\, \, \omega=\beta K^2\, .
\label{eq:Komega}
\end{equation}
The canonical form of the equations of motion is retrieved by rescaling time as $\tau=C_\tau t$, with
\begin{equation}
 C_\tau= \omega/K = \beta K = 2^{-2/3} \beta^{1/3} \gamma^{2/3} \, .
\label{eq:Ctau}
\end{equation}
We obtain $d R/d\tau=-\partial\Ham/\partial r$ and $d r/d\tau=\partial\Ham/\partial R$. Since \smash{$R'=\OO(m_j m_\star^{1/2} e_j^2)$} and \smash{$K=\OO(m_j^{5/3}/m_\star^{1/6})$}, the new action $R$ has order of magnitude \smash{$R=\OO\left(e_j^2\right)\OO\left(m_\star/m_j\right)^{2/3}$}. Defining the last parameter $\delta$ as
\begin{equation}\label{eq:delta}
\begin{aligned}
	\delta & =\alpha\left(\frac{4}{27\beta\gamma^2}\right)^{1/3}-1 \, \\
     & = -\frac{2}{3} \left( \frac{\Delta G}{ K} + S \right) -1 \, ,
    \end{aligned}
\end{equation}
the one-parameter and one-degree-of-freedom Hamiltonian of the SFMR of resonance reads
\begin{equation}\label{eq:SFM_polar}
	\Ham(R; r )=3\left(\delta+1\right)R -R^2+2\sqrt{2R}\cos r.
\end{equation}
Using the Cartesian canonical coordinates
\begin{equation}
	X=\sqrt{2 R }\cos r ,\;\;\;Y=\sqrt{ 2 R }\sin r,
\end{equation}
the Hamiltonian is written
\begin{equation}\label{eq:XY_1param}
	\Ham(X;Y)=\frac{3}{2}\left(\delta\!+\!1\right)\!\left(X^2\!+\!Y^2\right)\!-\!\frac{1}{4}\!\left(X^2\!+\!Y^2\right)^2\!+\!2X,
\end{equation}
or equivalently
\begin{equation}
	\Ham(X;Y)=2X-\frac{1}{4}\left(X^2+Y^2-3\left(\delta+1\right)\right)^2\, ,
\end{equation}
where constant terms are dropped. The equilibria are located at $Y=0$ and $X$ the roots of
\begin{equation}
	X^3-3\left(\delta+1\right)X-2=0.
\end{equation}
The equilibrium values of $X$ are then
\begin{equation}
	\begin{split}
		&X_\text{eq}\in\left\lbrace u_++u_-,\;\;ju_++j^2u_-,\;\;j^2u_++ju_-\right\rbrace\!,\;\text{where}\!\!\!\!\!\!\!\!\!\!\!\!\\
		&u_\pm=\left(1\pm\sqrt{1-\left(\delta+1\right)^3}\right)^{1/3},
	\end{split}
\end{equation}
and $j=\exp(2\iota\pi/3)$. With the constraint that $X_\text{eq}$ is a real number, there is a bifurcation at $\delta=0$. When $\delta<0$, only one (elliptic) fixed point exists, whereas for $\delta>0$, three fixed points exist, one of them being hyperbolic from which a separatrix emanates and defines the mathematical resonance (see Fig.~\ref{fig:SFM_anarea}).

\subsection{Definition of useful quantities}
%
%

We first define:
\begin{equation}
	\begin{split}
		M_C  = & \frac{m_1}{ m_\star} + \left( \frac{p}{p+1} \right)^{2/3} \frac{m_2}{m_\star} ,\\
		M^2  = & \left( \left(\frac{p}{p+1} \right)^{2/3} + \left(\frac{p}{p+1} \right)^{5/3} \frac{f_2^2}{f_1^2} \right)\frac{m_1 m_2}{m_\star^2} +\\
		 & \left(\frac{p}{p+1} \right)^{4/3} \left( \frac{m_2}{m_\star}\right)^2 + \frac{p}{p+1}  \frac{f_2^2}{f_1^2} \ \left( \frac{m_1}{m_\star}\right)^2.\\
	\end{split}
\label{eq:MMC}
\end{equation}
We note that for large $p$, $f_2/f_1 \approx \sqrt{(p+1)/p}$ \citep{DeBa2015}, so we have $M \approx M_C$. However, we still distinguish $M$ and $M_C$ to obtain the correct dependency in $m_1$ and $m_2$ for low $p$ values (in particular the 2:1 MMR). 

For the quantities of eq. (\ref{eq:abg}), we obtain: 
\begin{equation}
	\begin{split}
        &\beta=\frac{3}{2} \frac{p (p+1)n_{1,0} }{\sqrt{\G  a_{2,0}}} \frac{M_C \sqrt{m_\star}}{m_1m_2 },\\
        &\gamma=n_{1,0}|f_1| \sqrt{\frac{p}{p+1}}(\G a_{2,0})^{1/4}\frac{M \sqrt{m_1 m_2}}{M_C^{1/2 }m_\star^{1/4}}
	\end{split}
\end{equation}
while $C_\tau$ and $\delta$ become: 
\begin{equation}
	\begin{split}
	 C_\tau & = \frac{3^{1/3}}{2} n_1 \left( p |f_1| M\right)^{2/3} , \\
	\delta  & = \frac{2}{3} \left(  \frac{C'_\delta}{ M^{2/3}} \Delta_R + R \right) -1,
	\end{split}
	\label{eq:delCtau}
\end{equation}
where 
\begin{equation}
	\begin{split}
	C'_\delta & =\left( \frac{3 p ^{1/2}}{|f_1|} \right)^{2/3},\\
	\Delta_R &= \Delta \Lambda_1/ \Lambda_{1,0} -  \Delta \Lambda_2/ \Lambda_{2,0} = 1 - \left( \frac{p}{p+1} \frac{P_2}{P_1} \right)^{1/3}\, ,\\
    &\approx \frac{2\pi}{3 p n_{1,0}} \left( \frac{p+1}{P_2}-\frac{p}{P_1}\right) =\frac{2\pi}{3 p n_{1,0}} f_{super} \, .\\
	\end{split}
\end{equation}	
Thus, $\Delta_R$ is a parametrisation of the distance to the commensurability, and linked to the instantaneous super-period which evolves over the resonant timescale (we discussed the averaged super period in eq. \ref{eq:dvarphi2dt}). Defining
\begin{equation}
	\begin{split}
	C_\delta & =  \frac{2\pi}{3 p n_{1,0}} C'_\delta \, ,\\
	\end{split}
\end{equation}	
we can therefore write :
\begin{equation}
	\delta  \approx \frac{2}{3} \left(  \frac{C_\delta}{ M^{2/3}} f_{super} + R \right) -1\, .
	\label{eq:delfsup}
\end{equation}
Using equation (\ref{eq:delCtau}), and replacing $R$ by $\er^2/2$ (hence choosing the phase of the resonant mode), we obtain: 
%
\begin{equation}
\begin{aligned}
\Delta_R & = \frac{1}{2} ( 3 (\delta+1) - \er^2 )  M^{2/3}/C'_\delta\, ,\\
\end{aligned}
\label{eq:DR}
\end{equation}
We can therefore define
\begin{equation}
\begin{aligned}
\Delta_{\rm MMR} &= \er^2 - 3(\delta+1) \\
&= 2\left(\frac{3 p^{1/2}}{|f_1|}\right)^{2/3} M^{-2/3} \left[ \left(\frac{P_2}{P_1} \frac{p}{p+1} \right)^{1/3} -1 
\right], \\
 &\approx \frac{1 }{3^{1/3}\pi(p |f_1|)^{2/3}} M^{-2/3} \overline{P}_1 f_{super} = f_{super}/C_\tau \,.
\end{aligned}
\label{eq:DMMRap1}
\end{equation}
To better understand the scale of each MMR, we define $C_{p}$ which depends only on the considered $p$ resonance, and is defined as $C_p = M/\eps $ in the case $m_1=m_2$:
\begin{equation}
\begin{aligned}
		4 C_{p}^2 =  & \left(\frac{p}{p+1} \right)^{2/3} + \left(\frac{p}{p+1} \right)^{5/3} \frac{f_2^2}{f_1^2} \\  
        & + \left(\frac{p}{p+1} \right)^{4/3} + \frac{p}{p+1}  \frac{f_2^2}{f_1^2} \, . 
\end{aligned}
\label{eq:Cp}
\end{equation}
It can then be used in the general case by defining 
\begin{equation}
		M_p = M/C_p\, .
\label{eq:Mp}
\end{equation}
This gives : 
\begin{equation}
\begin{aligned}
\Delta_{\rm MMR} =K_p M_p^{-2/3} \left[ \left(\frac{P_2}{P_1} \frac{p}{p+1} \right)^{1/3} -1 
\right], 
\end{aligned}
\label{eq:DMMRap}
\end{equation}
where $\smash{K_p=2 (3 p^{1/2}/|f_1|)^{2/3} C_p^{-2/3}}$. We also define 
\begin{equation}
\begin{aligned}
\Delta_{p+1:p}  &= K_p  \left[ \left(\frac{P_2}{P_1} \frac{p}{p+1} \right)^{1/3} -1 
\right].
\end{aligned}
\label{eq:Dpp1pap}
\end{equation}
useful when the masses of the planets are unknown. 

\section{Validity of the model}
\label{ap:validity}

\begin{figure}
\begin{center}
\includegraphics[width=0.99\linewidth]{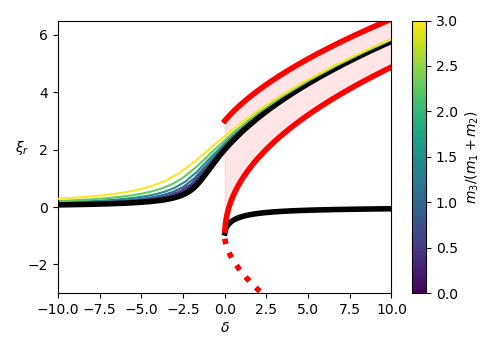}
\caption{Validity of the SFMR for planets of various masses embedded in a 3:2 - 3:2 resonant chain. We show in the SFMR the position of the fixed points of the chain.
%
}
\label{fig:3232} 
\end{center}
\end{figure}

\begin{figure}
\begin{center}
\includegraphics[width=0.99\linewidth]{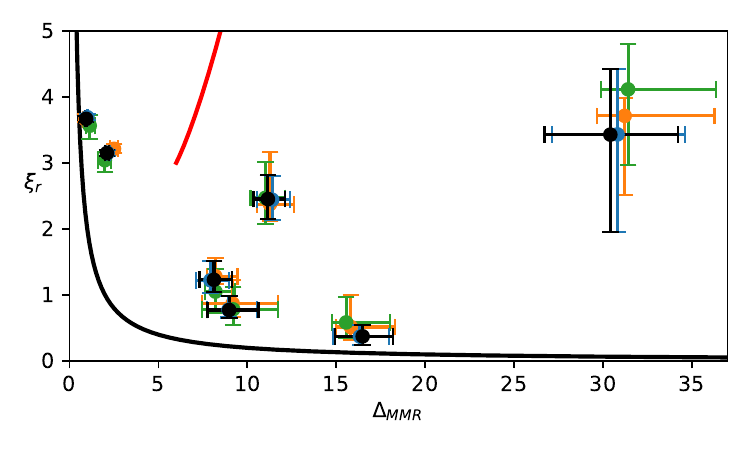}\\
\includegraphics[width=0.99\linewidth]{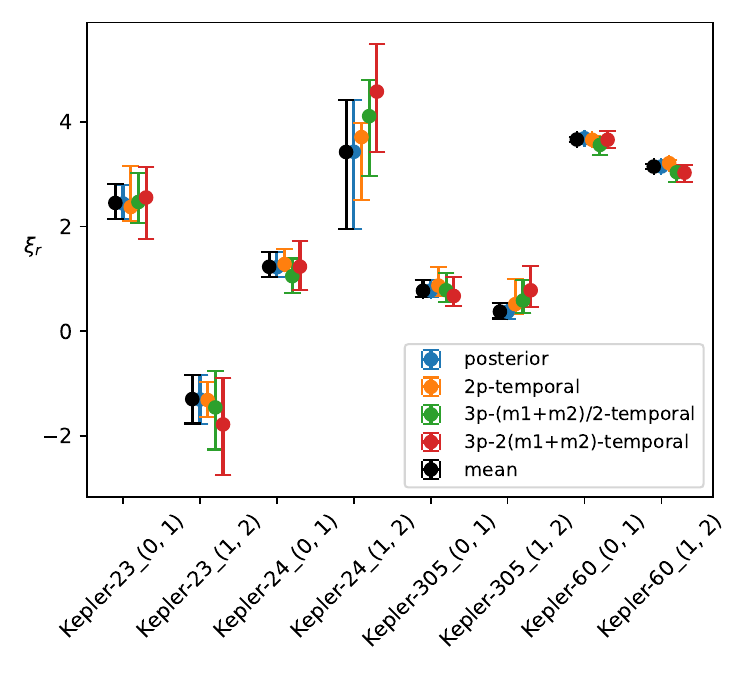}
\caption{Validity of the SFMR for observed pairs embedded in 3-planet systems. Blue shows direct projection of the TTV posterior of each pair, as used in this paper. Black shows the effect of going from osculating to mean orbital elements; orange shows the effect of the slow variations of the parameters assumed constant in the SFMR assuming an isolated pair. Green shows the effect of the addition of a planet whose mass is the average of the two interacting masses. Red shows (bottom only) the effect when the additional planet mass is twice the more massive of the two planets.
}
\label{fig:validity} 
\end{center}
\end{figure}

{We estimate how neighbouring planets could impact the representation of a given pair in the SFMR. In particular, we want to estimate how the presence of additional planets could increase the amplitude of libration $\Delta \er$ estimated for a given pair. To do so we chose a grid of masses, such that $m_2=10^{-5} m_\star$, $m_3/m_2 \in [0.1,10]$ and $m_2/m_1 \in [0.1,10]$. Then for each case, we compute the orbital elements of the families of elliptic fixed points of the chosen resonant chain (3:2-3:2 for Fig.~\ref{fig:3232})
using celeries\footnote{\url{https://gitlab.unige.ch/delisle/celeries/}} \citep{Delisle2017}. Then, for a grid of these fixed points, we compute where they land in the SFMR of the inner pair. The results, shown in Fig.  \ref{fig:3232}, show that the relevant parameter is the ratio between the `perturber' $m_3$, and the sum of the masses of the considered pair, $m_1+m_2$. In the top panel, we see that the maximum shift is reached for $\delta \sim -1$. We can see that the equilibrium point is shifted by $\sim 0.3$ when $m_3 \approx (m_1+m_2)$. On the other hand, when there is a non-consecutive first-order resonance in the chain, such as in a 3:2-4:3 configuration, the individual pairs are more affected \citep{Delisle2017}: to obtain a shift of 0.3 or less, we need $m_3 \lesssim (m_1+m_2)/2$. For \S\ref{sec:reso}, in which we focus on small $\Delta \er$ in the resonant domain, we filter the systems such that the maximum shift in $\Delta \er$ is below $0.15$. This is achieved by selecting $m_3 \lesssim (m_1+m_2)/2$ in the general case, and $m_3 \lesssim (m_1+m_2)/4$ in the case of non-neighbouring first-order MMR. }

Figure \ref{fig:validity} explores more generally how the various assumptions of the model (mean orbital elements, constant $\es$, isolated pairs of planets) impact the projection of a pair of planets onto the SFMR, on a subset of 8 pairs embedded in 3-planet systems. The blue markers represent what we do in the paper: the direct projection from the astrocentric osculating element from the database onto the SFMR. The black markers show the effect of averaging orbital elements. The orange markers are the result of a 1000-year numerical integration of 20 randomly selected samples of the posterior, for the considered pair only. The green and red markers show the same 1000 years of evolution, adding the third planet, with a mass either set at half the sum of the masses of the considered pair (green), or twice that sum (red). Although all these effects create some change in the projection, we consider that the aforementioned criteria ($m_3 \lesssim (m_1+m_2)/2$ in the general case, and $m_3 \lesssim (m_1+m_2)/4$ in the case of non-neighbouring first-order MMRs) are enough to ensure the applicability of the model.

\section{Temporal evolution of $e_j$ and $\varpi_j$}
\label{ap:ejwj}

Here we describe the evolution of $e_j$ and $\varpi_j$ for a pair of planets near, or into, a first order mean motion resonance, by inverting the changes of coordinates made in \S \ref{ap:Hamiltonian}. The evolution of the ($R',r$) degree of freedom is given by: 
\begin{equation}
\begin{aligned}
\frac{d r}{dt} = \frac{\partial \Ham'}{\partial R'} \, ; \, \, \frac{d R'}{dt} = -\frac{\partial \Ham'}{\partial r} \\
\end{aligned}
\end{equation}
hence
\begin{equation}
\begin{aligned}
\frac{d r}{dt} &= \alpha - 2 \beta R' + \frac{\gamma}{\sqrt{2R'}} \cos(r) \\
 \frac{d R'}{dt} &= \gamma \sqrt{2R'} \sin (r)
 \label{eq:eqRpr}
\end{aligned}
\end{equation}

Using eq. (\ref{eq:uv}) to (\ref{eq:UV}), we obtain:
\begin{equation}
\begin{aligned}
\sqrt{2D_1}e^{i\sigma_1} &= \cos(\phi) \sqrt{2R'}e^{i r} - \sin(\phi)  \sqrt{2S'}e^{i s} \, , \\
\sqrt{2D_2}e^{i\sigma_2} &=  \sin(\phi) \sqrt{2R'}e^{i r} + \cos(\phi)  \sqrt{2S'}e^{i s} \, .
\end{aligned}
\end{equation}
with:
Inversely, ($R'$,$r$) and ($S'$,$s$) write:
\begin{equation}
\label{eq:RpSp}
\begin{aligned}
R'&= \frac{f_1^2e_1^2 + f_2^2e_2^2 + 2 f_1f_2 e_1e_2 \cos (\varpi_1-\varpi_2)}{2(f_1^2/\Lambda_{1,0} + f_2^2/\Lambda_{2,0})}\\
S'&= \frac{ \frac{\Lambda_{1,0}}{\Lambda_{2,0}} f_2^2 e_1^2 +  \frac{\Lambda_{2,0}}{\Lambda_{1,0}} f_1^2 e_2^2 - 2 f_1f_2 e_1e_2 \cos (\varpi_1-\varpi_2)}{2(f_1^2/\Lambda_{1,0} + f_2^2/\Lambda_{2,0})}
\end{aligned}
\end{equation}
and 
\begin{equation}
\begin{aligned}
r&= - p \lambda_1 + (p+1) \lambda_2 - \varpi_r \\
s&= - p \lambda_1 + (p+1) \lambda_2 - \varpi_s \\
\end{aligned}
\end{equation}
with
\begin{equation}
\begin{aligned}
\varpi_r &= - \arg (f_1 e_1 \exp^{- \iota \varpi_1} + f_2 e_2 \exp^{- \iota \varpi_2}) \\
\varpi_s &= - \arg \left(-f_2 e_1 \sqrt{\frac{\Lambda_{1,0}}{\Lambda_{2,0}}} \exp^{- \iota \varpi_1} + f_1 e_2 \sqrt{\frac{\Lambda_{2,0}}{\Lambda_{1,0}}} \exp^{- \iota \varpi_2} \right)  \, .
\end{aligned}
\end{equation}

Since at first order of eccentricity $\sqrt{2D_j}=\sqrt{\Lambda_j}e_j$ and $\sigma_j=\varphi_2 - \varpi_j$, we obtain:
\begin{equation}
\begin{aligned}
e_1 e^{i\varpi_1} &= \sqrt{\frac{1}{\Lambda_1}} \left(\cos(\phi) \sqrt{2R'}e^{i ( \varphi_2 -r)}- \sin(\phi)  \sqrt{2S'}e^{i (\varphi_2 -s)}\right) , \\
e_2 e^{i\varpi_2} &=   \sqrt{\frac{1}{\Lambda_2}} \left(\sin(\phi) \sqrt{2R'}e^{i ( \varphi_2 -r)}+ \cos(\phi)  \sqrt{2S'}e^{i (\varphi_2 - s)}\right) .
\label{eq:evarpi}
\end{aligned}
\end{equation}
which we rewrite as:
\begin{equation}
\begin{aligned}
e_1 e^{i\varpi_1} &= e_{r,1}(t) \e^{i \varpi_r(t)}
+  e_{s,1} \e^{i \varpi_s}\, , \\
e_2 e^{i\varpi_2} &= - e_{r,2}(t) \e^{i \varpi_r(t)}
+  e_{s,2} \e^{i \varpi_s},
\label{eq:evarpin}
\end{aligned}
\end{equation}
with:
\begin{equation}
\begin{aligned}
e_{r,1}(t) &=  \sqrt{\frac{1}{\Lambda_1}} \cos(\phi) \sqrt{2R'(t)}  \, , \\  
e_{r,2}(t) &= -\sqrt{\frac{1}{\Lambda_2}} \sin(\phi) \sqrt{2R'(t)}\, ,\\
e_{s,1}(t) &= - \sqrt{\frac{1}{\Lambda_1}} \sin(\phi) \sqrt{2S'} \, ,\\ 
e_{s,2}(t) &= \sqrt{\frac{1}{\Lambda_2}} \cos(\phi) \sqrt{2S'}\,,
\label{eq:erjwj}
\end{aligned}
\end{equation}
and $\varpi_r(t)= \varphi_2-r$. Since $\cos \phi>0$ and $\sin \phi<0$ the definitions \ref{eq:erjwj} ensure positive eccentricities, while eq (\ref{eq:evarpin}) shows that the resonant contributions are anti-aligned, while the secular ones are aligned.
The evolution of the angles $\varphi_2$ and $s$, associated with the conserved momenta $\Delta G$ and $S'$ respectively, is described by:
\begin{equation}
\begin{aligned}
\frac{d \varphi_2}{dt} = \frac{\partial \Ham'}{\partial \Delta G} \, ; \, \, \frac{d s}{dt} = \frac{\partial \Ham'}{\partial S} \, .
\end{aligned}
\end{equation}
Interestingly, these derivatives are identical in our model:
\begin{equation}
\label{eq:dvarphi2dt}
\begin{aligned}
\frac{d \varphi_2}{dt} &=\frac{d s}{dt} = \alpha - 2\beta R'= -C_\tau \Delta_{\rm MMR} = -f_{super}
\end{aligned}
\end{equation}
This implies that, within this model, the contributions of $S'$ to the eccentricity vectors (second terms in eq. \ref{eq:evarpi}) are constant. The full dynamics of ($R'$,$r$) is described by the Hamiltonian eq. (\ref{eq:secondFundamentalModel}) and will be described in \S\ref{sec:1parameter}. Here we consider two cases: small oscillations in the resonant domain and the case far from resonance.

\subsection{Small resonant oscillations}
\label{ap:smallosc}
As seen in \S\ref{sec:1parameter}, one can define the elliptic equilibrium ($R'=R'_0$,$r=0$) and expand equation (\ref{eq:eqRpr}) in its neighborhood. Considering small variations around that equilibrium and introducing $\tilde r =r$ and $\tilde R' = R'-R'_0$, we obtain:
\begin{equation}
\begin{aligned}
\frac{d \tilde r}{dt} &=- \left(  2 \beta + \frac{\gamma}{(2R'_0)^{3/2}} \right) \tilde R' = -A \tilde R' \\
 \frac{d \tilde R'}{dt} &= \gamma \sqrt{2R'_0} \tilde r = B \tilde r
 \label{eq:eqRpr}
\end{aligned}
\end{equation}
whose solution is that of a harmonic oscillator of frequency $\omega=\sqrt{AB}$. Taking, for example, $\tilde R'(0)=\tilde R'_i$ and $r(0)=0$, we obtain:
\begin{equation}
\begin{aligned}
\tilde r(t)  &= -\tilde R'_i \sqrt{\frac{A}{B}} \sin (\omega t)\\
\tilde R'(t) &= \tilde R'_i \cos (\omega t)\\
 \label{eq:eqRpr}
\end{aligned}
\end{equation}


The evolution of $\varphi_2$ is given by:
\begin{equation}
\label{eq:dvarphi2dt}
\begin{aligned}
\frac{d \varphi_2}{dt} &=  - 2 \beta (\Delta G +R'_0 +S')   - 2 \beta \tilde R'\\
&= 3 n_{1,0} p \left(  1 - \left( \frac{p}{p+1}\frac{\overline{P}_2}{\overline{P}_1} \right)^{1/3}\right)  - 2 \beta \tilde R'\\
&= -2 \pi \left( \frac{p}{\overline{P}_1} - \frac{p+1}{\overline{P}_2} \right)  - 2 \beta \tilde R' + \gO\left( \left( \frac{p}{\overline{P}_1} - \frac{p+1}{\overline{P}_2} \right)^2 \right)\\
&\approx -\overline{f}_{super} - 2 \beta \tilde R' \, .
\end{aligned}
\end{equation}
where $\overline{P}_j$ is the period of planet $j$ at the equilibrium, which is equivalent to the period averaged over the resonant motion in our small oscillation approximation. The first term of $d \varphi_2/dt$ is, at first order, the frequency associated with the super-period \citep[e.g.][]{Lithwick2012}. 

$R',r$ evolve over the resonant timescale as described in \S\ref{sec:1parameter}. In our small oscillation approximation we obtain:
\begin{equation}
\frac{d \varpi_r(t)}{dt} =  - \overline{f}_{super} + \frac{\gamma}{(2R'_0)^{3/2}} \tilde R'_i \cos(\omega t)
\end{equation}

\subsection{Far from resonance }
\label{ap:far}
We now consider the case far from resonance, with $| \alpha - 2\beta R'|=|f_{super}|\!>\!>\!\gamma/\sqrt{2R'}$. At $0^{th}$ order in $\gamma$, hence the planetary masses, we obtain:
\begin{equation}
\begin{aligned}
\frac{d  r}{dt} &= \alpha - 2\beta R' +\gO(\gamma)= -f_{super} +\gO(\gamma)\, , \\
 \frac{d  R'}{dt} &= \gO(\gamma)\, ,
 \label{eq:eqRprfarres}
\end{aligned}
\end{equation}
where $f_{super}$ is evaluated at $R_0$. The angle $r$ therefore circulates and $R'=R'_0$ is constant.
At first order in $\gamma$, we obtain:
\begin{equation}
\begin{aligned}
R'(t) &=R_0' + \frac{\gamma \sqrt{2R'_0}}{f_{super}} \cos(-f_{super}t +r_0)\, .
\label{eq:eqRprfarres}
\end{aligned}
\end{equation}
Regarding the evolution of $\varpi_r$, we have:
\begin{equation}
\begin{aligned}
\frac{d \varpi_r(t)}{dt} &=  \frac{d (\varphi_2-r)}{dt}   \\
&=-\frac{\gamma}{\sqrt{2R_0'}}\cos(-f_{super}t+r_0)
\end{aligned}
\label{eq:R_farres}
\end{equation}
hence
 \begin{equation}
\begin{aligned}
\varpi_r(t) &= \varpi_r(0) +\frac{\gamma}{\sqrt{2R'_0}C_\tau \Delta_{\rm MMR}}\sin(-f_{super} t+r_0)\, .
\end{aligned}
\label{eq:pomegar_farres}
\end{equation}
We can also compute $h_j=e_j \sin \varpi_j$. Noting $\varphi_{2,0} = r_0 + \varpi_{r,0}$ the phase of the line of conjunction $\varphi_2= -p n_1t + (p+1) n_2  t  +  \varphi_{2,0}$, we obtain:
 \begin{equation}
\begin{aligned}
h_1&= \sqrt{\frac{1}{\Lambda_1}} \cos \phi \frac{\gamma} {f_{super}} \sin (-f_{super}t + \varphi_{2,0})\, ,\\
h_2&= \sqrt{\frac{1}{\Lambda_2}} \sin \phi \frac{\gamma} {f_{super}} \sin (-f_{super}t + \varphi_{2,0})\, .
\end{aligned}
\label{eq:pomegar_farres}
\end{equation}

\section{Degeneracies of the TTV signal}
\label{ap:degen}

\begin{figure}
\begin{center}
\includegraphics[width=0.99\linewidth]{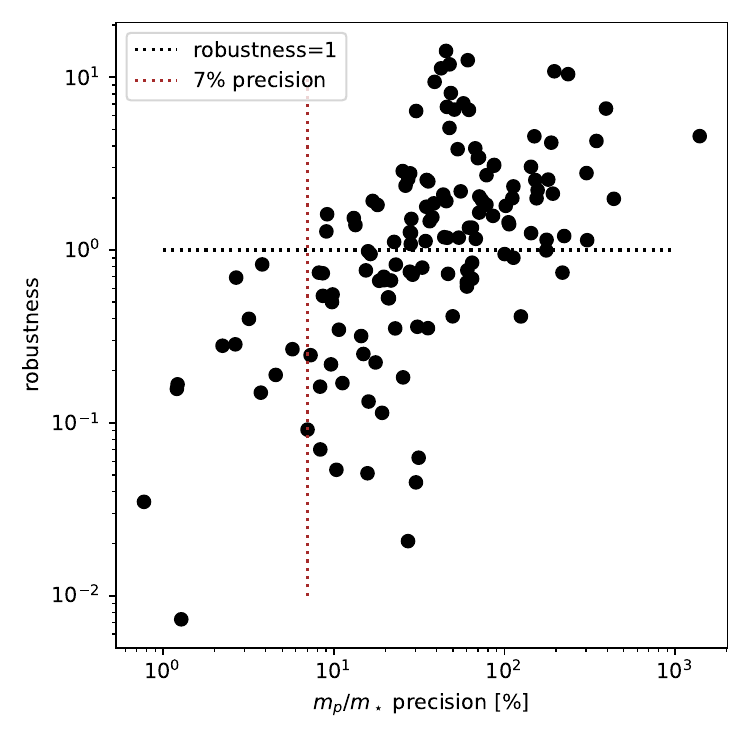}
\caption{Mass precision vs robustness using posteriors from \cite{HaLi2017} and \cite{Leleu2023}. The mass precision is used on the `default' set of priors: log-uniform in mass and uniform in eccentricities. The robustness criterion defined in \cite{HaLi2017} is below 1 if the median of the posterior using the `highmass' (uniform in mass and log-uniform in eccentricities) lies within the 0.16-0.84 quantiles of the `default' set of priors, and above 1 if it lies outside.
}
\label{fig:obs_rob} 
\end{center}
\end{figure}

\begin{figure}
\begin{center}
\includegraphics[width=0.99\linewidth]{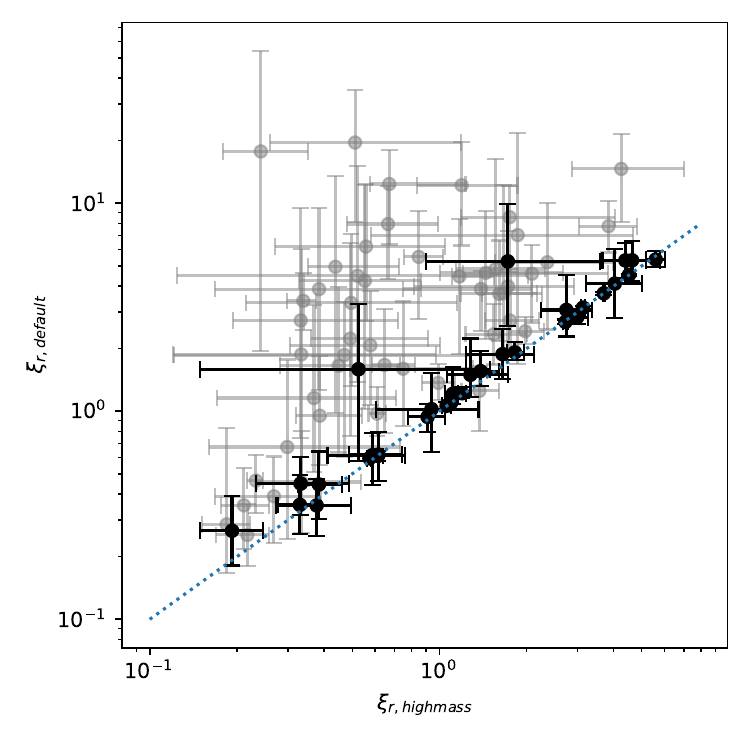}
\includegraphics[width=0.99\linewidth]{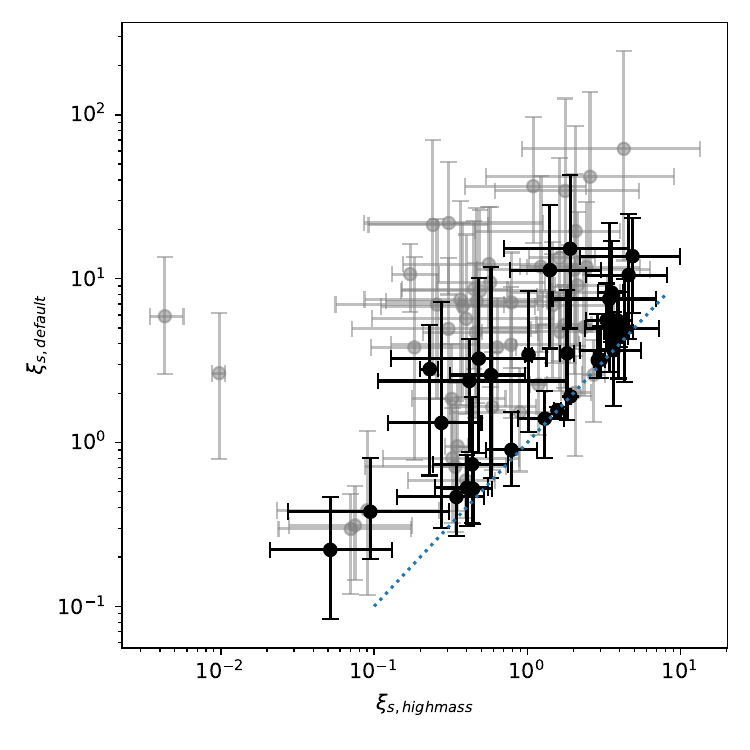}
\caption{Comparison of $\er$ and $\es$ \textit{default} and \textit{highmass} posteriors from \cite{HaLi2017} and \cite{Leleu2023}. A parameter is deemed robustly estimated if the two sets of priors give similar results.
}
\label{fig:obs_rob} 
\end{center}
\end{figure}

\subsection{Expression of the TTVs}


{We adapt the model of analytical TTVs from \cite{NeVo2016} to the formulation of the SFMR described in Appendix \ref{ap:Hamiltonian}. The TTVs of the planet $j$ can be approximated by \citep{Nesvorny2009}:}

\begin{equation}
\delta_{t_j} = \frac{1}{n_j} (-\delta_{\lambda_j} + 2 \delta_{h_j} ) + \gO(e_j),
\label{eq:deltatj}
\end{equation}
{where $n_j$ is the mean motion of planet $j$ and $h_j= e_j \sin \varpi_j$. As described in sections 2 and 4 of \cite{NeVo2016}, near and close to the resonance, the term $\delta_{\lambda_j}$ is the main contribution to the TTVs at the resonant frequency, while $2\delta_{h_j}$ mainly contributes over a slower time scale. As we are mainly interested in the degeneracy that affects the resonant dynamics, we develop a simple sinusoidal TTV model based on the contribution of $\delta_{\lambda_j}$. We will check later for the contribution of $2\delta_{h_j}$ far from the resonance. From equation \ref{eq:HamKep}, we obtain:}
\begin{equation}
\dot \lambda_j = n_{j,0} - 3 n_{j,0} \frac{\Delta \Lambda_j}{\Lambda_{j,0}},
\label{eq:dotlambdaj}
\end{equation}
From eq. (\ref{eq:DG}), we obtain:
\begin{equation}
\dot \lambda_1 = C_{\lambda_1} + 3 p \frac{n_{1,0}}{ \Lambda_{1,0}}  R'  \, ,\ \ \dot \lambda_2 = C_{\lambda_2} - 3 (p+1) \frac{n_{2,0}}{  \Lambda_{2,0}} R',
\label{eq:dotlambdaj}
\end{equation}
{where $C_{\lambda_1}$ and $C_{\lambda_2}$ are constants. These constants will affect the average periods of the planets, but do not generate TTVs, so we do not need their explicit values. The temporal evolution of $\lambda_j (t)$ is therefore given by:}

\begin{equation}
\begin{aligned}
 \lambda_1 (t) &= C_{\lambda_1}  (t-t_0)+ 3 p \frac{n_{1,0}}{ \Lambda_{1,0}}  \int^t_{t_0} R' dt, \\
 \lambda_2 (t) &= C_{\lambda_2} (t-t_0) - 3 (p+1) \frac{n_{2,0}}{  \Lambda_{2,0}}   \int^t_{t_0} R' dt.
\label{eq:lambdajt}
\end{aligned}
\end{equation}

Using the rescaling of $R'$ and the time, we have:
\begin{equation}
\begin{aligned}
 \int^t_{t_0} R'(t) dt= \frac{1}{\beta} \int^\tau_{\tau_0} R(\tau) d\tau.
\label{eq:intR}
\end{aligned}
\end{equation}
{The exact analytical solution of the 1 degree of freedom Hamiltonian (\ref{eq:SFM_polar}) is given by \citet[][\S 3.6]{NeVo2016}. Here we derive an analog of their equations (28) and (29) using our Hamiltonian \ref{eq:SFM_polar} to replace their $a_j$ coefficients:}
\begin{equation}
\begin{aligned}
    a_0 &= -1\, , \\
    a_1 &= \frac{3}{2} (\delta+1) \, , \\ 
    a_2 &= - \left(\frac{3}{2}  (\delta+1)^2 + \frac{h_0}{3} \right) \, , \\
    a_3 &= 2 +  \frac{3}{2}  (\delta+1) h_0   \, , \\
    a_4 &= - h_0^2 \, , \\
\end{aligned}
\end{equation}
{where $h_0$ is the value of the Hamiltonian, eqn.\ \ref{eq:SFM_polar}, at epoch. \cite{NeVo2016} note that the momentum $R$ follows }
\begin{equation}
\left( \frac{d R}{d \tau} \right)^2 =  a_0 R^4 + 4 a_1 R^3 + 6 a_2 R^2 + 4 a_3 R + a_4,
\label{eq:dR2}
\end{equation}
{whose general solution is a function of Weierstrass elliptic functions. The first term of the Fourier expansion of the solution can be written as (NV16 eq.59):}
\begin{equation}
 R(\tau)= R_0+D(\delta,\er) - D(\delta,\er) \cos(f_\tau \tau + \varphi_R),
\label{eq:Rtau}
\end{equation}
{where $D(\delta,\er)$ is the semi-amplitude of the oscillation of the action $R$ in the SFMR (\ref{eq:SFM_polar}), and $f_\tau(\delta,\er)$ is its frequency. The variables $D$, $f_\tau$ and the phase $\varphi_R$ depend on the initial value of the resonant mode ($R$,$r$). The phase of the resonant mode will be discussed in appendix \ref{ap:phases}. Here, as we are only interested in the amplitude and period of the TTVs, we replace $R$ by its largest value within the orbit: $\er^2/2$. Then, $D(\delta,\er)$ and $f_\tau(\delta,\er)$ can be estimated analytically by following \S 3.7 and \S 4.1 of NV16. For $D(\delta,\er)$, we evaluate it semi-analytically by following the iso-energy curves of the Hamiltonian (\ref{eq:SFM_polar}) and computing half of $\max(R) -\min(R)$ along the curve (these extrema are reached for $r=0$ and $\pi$).} 

{To estimate the TTVs of each planet, we substitute $t$ for $\tau$ using eqn.\ \ref{eq:Ctau}, then we substitute eqn.\ \ref{eq:Rtau} into (\ref{eq:lambdajt}) and keep only the periodic terms: }
\begin{equation}
\begin{aligned}
\delta_{t_1} = \frac{\delta_{\lambda_1}}{n_{1,0}} &= - 3 p   \frac{1}{ \Lambda_{1,0}} \frac{D(\delta,\er)}{\beta} \frac{P_\tau(\delta,\er)}{2\pi} \sin(f t), \\
\delta_{t_2} =\frac{\delta_{\lambda_2}}{n_{2,0}} &=  3 (p+1)  \frac{1}{  \Lambda_{2,0}}   \frac{D(\delta,\er)}{\beta} \frac{P_\tau(\delta,\er)}{2\pi} \sin(f t), 
\label{eq:TTVana}
\end{aligned}
\end{equation}
where $P_\tau(\delta,\er) = 2\pi /f_\tau$ is the period of the resonant mode for the SFMR (\ref{eq:SFM_polar}).

We now focus on the other two observables $A_{TTV}$ and $P_{TTV}$. Recalling the expression of $\beta$ (eq. \ref{eq:abg}), 
%
it appears that the sum of the TTV amplitudes 
\begin{equation}
\begin{aligned}
A_{TTV} & =  \frac{D(\delta,\er) P_\tau(\delta,\er)}{ p n_{1,0} \pi}\, , \\
\end{aligned}
\label{eq:ATTV}
\end{equation}
does not depend on the planetary masses. The period of the TTVs is the resonant period: $P_{TTV}  = P_\tau(\delta,\er)/C_\tau$. Our ability to constrain three parameters ($\delta$,$\er$ and $M$) with three observables ($\Delta_R$,$A_{TTV}$ and $P_{TTV}$) depends on how independent these constraints are. This can be estimated by studying the determinant of the normalised Jacobian matrix $J$:
\begin{equation}
\begin{aligned}
J & = \begin{pmatrix}
   \frac{\partial \Delta_{R}}{\partial \er} , &  \frac{\partial A_{TTV}}{\partial \er} , & \frac{\partial P_{TTV}}{\partial \er} \\
      \frac{\partial \Delta_{R}}{\partial \delta} ,&  \frac{\partial A_{TTV}}{\partial \delta} , & \frac{\partial P_{TTV}}{\partial\delta} \\
         \frac{\partial \Delta_{R}}{\partial M} , & \frac{\partial A_{TTV}}{\partial M} ,&  \frac{\partial P_{TTV}}{\partial M} \\
  \end{pmatrix},\\
\end{aligned}
\label{eq:observables}
\end{equation}

The lower $\text{det}(J)$, the stronger the degeneracy: $\text{det}(J)=1$ implies fully orthogonal constraints, while $\text{det}(J)=0$ implies that the constraints are not linearly independent. 
To visualise the level of degeneracy in a 2D plane, we chose to fix the period ratio between the two planets, hence $\Delta_{p+1:p}$, which is usually the best constrained of the observables. This allows us to map values of the parameter $M$ for each value of ($\delta$,$\er$) by inverting eq. (\ref{eq:DR}).

As can be seen in Fig.~\ref{fig:SFM_degen} and \ref{fig:SFM_degen}, the resonance-trapped domain is the least affected by the degeneracy, while the degeneracy is stronger far from the resonance and larger values of $|\er|$. This corresponds to the part of the phase space where $P_{TTV}$ gets close to the super period.

We now consider the effect of $\delta_{2h_j}$ in the TTV signal eq. (\ref{eq:deltatj}). Inside the resonance, our small oscillation approximation \S\ref{ap:smallosc} shows that the $2h_j$ contribution is dominated by the precession of $\vec{e}_{r,j}$ at the super period, see Fig. \ref{fig:elements}. This adds a slower sinusoidal signal to the TTVs, visible on the bottom left panel of the figure.

We now compare the contributions of $\delta_{\lambda_j}$ and $\delta_{2h_j}$ to the TTVs far from resonance. Here we will keep track of the phases of each signal because we will use them in Appendix \ref{ap:phases}. Following from \S \ref{ap:far}, both $\delta_{\lambda_j}$ and $\delta_{2h_j}$ evolve over the super period:
\begin{equation}
\begin{aligned}
\delta_{\lambda_1} &=-3p\frac{n_{1,0}}{\Lambda_{1,0}} \gamma \frac{\sqrt{2R'_0}}{f_{super}^2} \sin (-f_{super} t + r_0)\\
\delta_{2h_1} &= \frac{2}{\sqrt{\Lambda_{1,0}}} \cos \phi \frac{\gamma}{f_{super}} \sin (-f_{super} t + \varphi_{2,0})
\end{aligned}
\label{eq:TTVfares}
\end{equation}
Computing the ratio of the two quantities (recalling $f_{super} = C_\tau \Delta_{\rm MMR}$, $\sqrt{2R_0'}= \sqrt{K} \er$, and using the expressions of $\beta$ and $\gamma$), we obtain:
\begin{equation}
\begin{aligned}
F_1=\frac{\delta_{\lambda_1}}{\delta_{2h_1}} &=\frac{\er}{\Delta_{\rm MMR}} \frac{M}{M_C}\left( \frac{3}{p|f_1|M}\right)^{1/3}.
\end{aligned}
\label{eq:F1ap}
\end{equation}
This ratio therefore depends on the position in the SFMR, but also on the considered resonance and planetary masses (to the power 1/3), and is akin to the ratio between forced and free eccentricities discussed in \cite{Lithwick2012}. 
The effect of $\delta_{2h_j}$ on the TTVs is therefore stronger at large $|\Delta_{\rm MMR}|$ and low $|\er|$. In that area, $\delta_{2h_j}$ adds to the degeneracy described when studying $\delta_{\lambda_j}$, and the solution can depart from the iso-amplitude curves described by eq. (\ref{eq:ATTV}). However, when studying the position in the iso-amplitude curve when far from resonance, we obtain:
\begin{equation}
\begin{aligned}
A_{TTV}=\frac{4 \er}{n_{1,0} p \Delta_{\rm MMR}^2}\, .
\end{aligned}
\label{eq:TTVfares}
\end{equation}
As $\er$ is proportional to $\Delta_{\rm MMR}^2$ along the iso-amplitude curve, eventually $\delta_{\lambda_j}$ dominates the TTV signal and the posterior is expected to follow that curve. 

\section{TTV Phases}
\label{ap:phases}



\begin{figure}
\begin{center}
\includegraphics[width=0.99\linewidth]{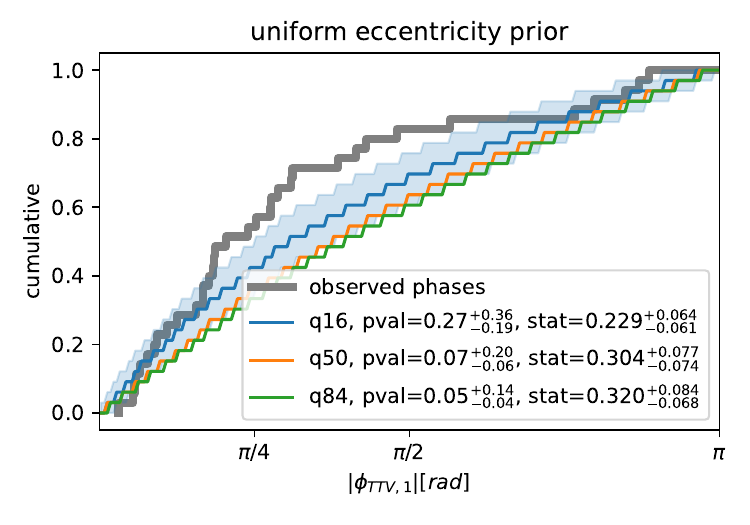}
\includegraphics[width=0.99\linewidth]{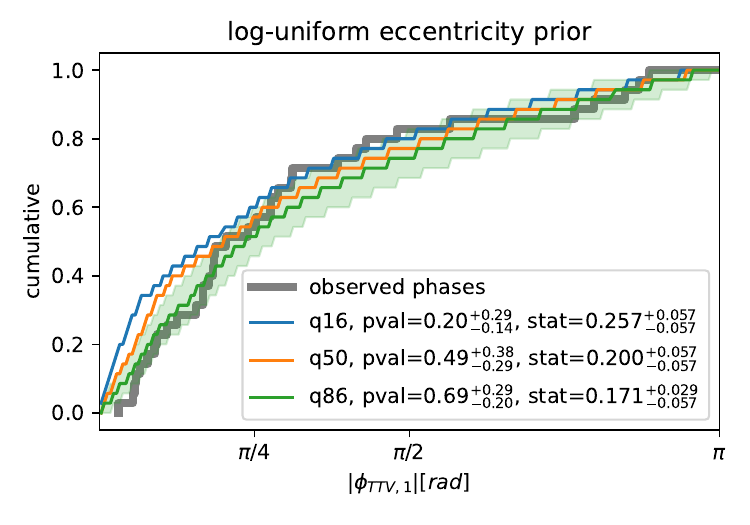}
\caption{
The grey curve shows the cumulative of $|\phi_{TTV,1}|$ for the systems in the upper circulation area. The coloured cumulatives show the phase distributions that are expected to be observed assuming that the true distribution of $\Delta \er$ is either the 0.16, 0.50 or 0.84 quantile of either the log-uniform or uniform eccentricity prior. For each case, we compute a Kolmogorov-Smirnov test. Taking as $\Delta \er$ distribution the 0.84 quantile of the log-uniform eccentricity prior gives the closest phase distribution. See the text for more details.
}
\label{fig:phase_cumul} 
\end{center}
\end{figure}

Another constraint that can be derived directly from the sinusoidal approximation of the TTVs is the TTV phase shift, $\phi_{TTV,j}$. The quantity $\phi_{TTV,j}$ was introduced by \cite{Lithwick2012} as the phase difference between the longitude of conjunction of the planets $\varphi_{2,0}$ and the phase at which the TTV signal of planet $j$ is equal to 0 while decreasing with time. \cite{Lithwick2012} showed that it potentially contains information on how far the pair of planets is from the elliptic fixed points. 

Indeed, for planets far from commensurability (in the sense of \S\ref{ap:far}), if the pair is at the fixed point, $\delta_{\lambda_1}=0$ in Eq.\ \ref{eq:TTVfares}. Therefore, the TTV signals come from $\delta_{2h_1}$ only ($F_1=0$), with its phase being the phase of the line of conjunction $\varphi_{2,0}$. As $R_0'$, hence $\er$ increases, $\delta_{\lambda_1}$ adds another sinusoidal term to the TTVs signal, with the same frequency but a different phase $r_0$ that is random. As long as $F_1<1$, $|\phi_{TTV,j}|$ has a maximum value smaller than $\pi/2$. Then, for $F_1>1$, the maximum value is $\pi$ (see Fig. \ref{fig:phase_distrib}). 
%

The approach of \cite{Lithwick2012} is only valid for pairs of planets. To generalize this to systems with 3+ planets, we recompute the TTV signal that would be produced by each pair of planets over 4 years, and estimate the $\phi_{TTV,1}$ values sample by sample, estimating their median values and error bars.

For a given pair, the information contained in $|\phi_{TTV}|$ with respect to the position in the SFMR is the minimum value of $\Delta \er$ that can reproduce the observed $|\phi_{TTV}|$. For $|\phi_{TTV}| \sim 0$, the pair can be anywhere, while for non-zero phase, there is a lower limit on the possible values of $\Delta \er$, and hence $\er$. 
However, we note that for a given position in the phase space, the probability for the pair to have a given phase within the available range is not uniform (see Fig.~\ref{fig:phase_distrib}). The phase distribution therefore contains information at the population level: for example, it is unlikely that all pairs have very large $\er$, because otherwise they would all have flat probability for $|\phi_{TTV}|$, and this is not what we observe (Fig. \ref{fig:phase_cumul}). 

To leverage this information, we assume different $\er$ distributions that are consistent with the data (16th, 50th, 84th percentile of a given set of priors, as well as different sets of priors) that we call a `case'. Then we generate for each case ten thousand possible distributions of phases, each following the $|\phi_{TTV,1}|$ probability distribution expected for each pair. This gives us, for each case, a population-level phase distribution for which we can compute, phase bin per phase bin, a median probability, as well as errorbar, examples of which are shown in Fig.~\ref{fig:phase_cumul}. Using these, we can verify which case best matches the observed phase distribution using a Kolmogorov-Smirnov test. The 0.84 quantile of the log-uniform eccentricity prior gives the highest p-value; we therefore choose the distribution of $\Delta \er$ for that case with the black dashed line of Fig. \ref{fig:Der_upper}.


\end{document}